\input ./style/arxiv-general.cfg
\documentclass[MSNbibl,nameyear,seceqn,dvips]{arxstspdf}
\makeatletter
   \@ifpackageloaded{graphicx}{}{\usepackage{graphicx}}
\makeatother
\usepackage{stfloats}

\volume{30}
\issue{4}
\pubyear{2015}
\firstpage{533}
\lastpage{558}
\doi{10.1214/15-STS527}% Updated by VTEXPTS2LaTeX.exe, 04.09.2015 10:46
\docsubty{FLA}

\makeatletter
\newcommand{\rrvert}{\vert}
\newcommand{\llvert}{\vert}
\def\cal{\mathcal}

\renewcommand{\citep}[1]{\citeauthor{#1}, \citeyear{#1}}
\newcommand{\PP}{\mathbb{P}}
\newcommand{\EE}{\mathbb{E}}
\newcommand{\eps}{\varepsilon}
\newcommand{\Cov}{\operatorname{Cov}}
\newcommand{\Var}{\operatorname{Var}}
\newcommand{\bx}{\mathbf{X}}
\newcommand{\argmin}{\operatorname{argmin}}

\newtheorem{theo}{Fact}

\newtheorem{lemm}{Lemma}

\newproclaim{defi}{Definition}
\newproclaim{examp}{Example}
\def\textsf{\texttt}
\makeatother

\begin{document}
\begin{frontmatter}

\title{High-Dimensional Inference:
Confidence Intervals, $p$-Values and \texttt{R}-Software \texttt{hdi}}
\runtitle{High-Dimensional Inference:
Confidence Intervals, $p$-Values and \textsf{R}-Software \texttt{hdi}}
%\title{}%\thanksref{T1}
% kai straipsnis turi susijusiu diskusiju ir rejoinder'iu
%\relateddois{T1}{Discussed in \relateddoi{d}{10.1214/00-STSXXX} ...;
%rejoinder at \relateddoi{r}{10.1214/00-STSXXXX}.}
%\runtitle{}
%\pdftitle{}

\begin{aug}
% Corresponding author: Ruben Dezeure - dezeure@stat.math.ethz.ch% Updated by VTEXPTS2LaTeX.exe, 07.09.2015 08:19
%Updated by VTEXPTS2LaTeX.exe, 04.09.2015 10:46
\author[A]{\fnms{Ruben}~\snm{Dezeure}\corref{}\ead[label=e1]{dezeure@stat.math.ethz.ch}},
\author[A]{\fnms{Peter}~\snm{B\"uhlmann}\ead[label=e2]{buhlmann@stat.math.ethz.ch}},
\author[A]{\fnms{Lukas}~\snm{Meier}\ead[label=e3]{meier@stat.math.ethz.ch}}
\and
\author[A]{\fnms{Nicolai}~\snm{Meinshausen}\ead[label=e4]{meinshausen@stat.math.ethz.ch}}

% \thankstext{t2}{Footnote to the first author with the `thankstext'
%command.}

\runauthor{Dezeure, B\"uhlmann, Meier and Meinshausen}

\address[A]{Ruben Dezeure is a Ph.D. student, Peter B\"uhlmann is Professor,
Lukas Meier is Senior Scientist and Nicolai Meinshausen is Professor,
Seminar for Statistics, ETH Z\"{u}rich, CH-8092 Z\"{u}rich, Switzerland \printead{e1,e2}, \printead*{e3,e4}.}

%\runauthor{}
%\pdfauthor{}
\end{aug}

% ABSTRACT

\begin{abstract}
We present a (selective) review of recent frequentist high-dimensional
inference methods for constructing $p$-values and confidence intervals in
linear and generalized linear models. We include a broad, comparative
empirical study which complements the viewpoint from statistical
methodology and theory. Furthermore, we introduce and illustrate the
\texttt{R}-package \texttt{hdi} which easily allows the use of different
methods and supports reproducibility.
\end{abstract}

% KEYWORDS

\begin{keyword}
\kwd{Clustering}
\kwd{confidence interval}
\kwd{generalized linear model}
\kwd{high-dimensional statistical inference}
\kwd{linear model}
\kwd{multiple testing}
\kwd{$p$-value}
\kwd{\texttt{R}-software}
\end{keyword}
%
%\begin{keyword}
%\kwd{}
%\end{keyword}
\end{frontmatter}
%
%s1 #&#
\section{Introduction}\label{sec1}

Over the last 15 years, a lot of progress has been achieved in
high-dimensional statistics where the number of parameters can be much
larger than sample size, covering (nearly) optimal point estimation,
efficient computation and applications in many different areas; see, for
example, the books by \citet{hastetal09}, \citet{pbvdg11} or the review
article by \citet{fanlv10}. The core task of statistical inference
accounting for uncertainty, in terms of frequentist confidence intervals
and hypothesis testing, is much less developed. Recently, a few methods for
assigning $p$-values and constructing confidence intervals have been
suggested
(\citep{WR08};
\citep{memepb09};
\citep{pb13};
\citep{zhangzhang11};
\citep{covtest14};
\citep{vdgetal13};
\citep{jamo13b};
\citep{meins13}).

The current paper has three main pillars: (i) a (selective) review of the
development in frequentist high-dimensional inference methods for $p$-values
and confidence regions; (ii) presenting the first broad, comparative
empirical study among
different methods, mainly for linear models: since the methods are
mathematically justified under
noncheckable and sometimes noncomparable assumptions, a thorough
simulation study should lead to additional insights about reliability and
performance of various procedures; (iii) presenting the \texttt{R}-package
\texttt{hdi} (\emph{h}igh-\emph{d}imensional \emph{i}nference) which
enables to easily use many of the different methods for inference in
high-dimensional generalized linear models. In addition, we include a
recent line of methodology allowing to detect significant groups of
highly correlated variables which could not be inferred as individually
significant single variables (\cite{meins13}). The review and exposition in
\citet{bumeka13} is vaguely related to points (i) and (iii) above, but
much more
focusing on an application oriented viewpoint and covering much less statistical
methodology, theory and computational details.

Our comparative study, point (ii), mentioned above, exhibits interesting
results indicating that more ``stable'' procedures based on
Ridge-estimation or random sample splitting with subsequent
aggregation are somewhat
more reliable for type I error control than asymptotically power-optimal
methods. Such results cannot be obtained by comparing underlying
assumptions of different methods, since these assumptions are often too
crude and far from necessary. As expected, we are unable to
pinpoint to a method which is (nearly) best in all considered
scenarios. In view of this, we also want to offer a collection of useful
methods for the community, in terms of our \textrm{R}-package \texttt{hdi}
mentioned in point (iii) above.

%s2 #&#
\section{Inference for Linear Models}\label{sec.LM}

We consider first a high-dimensional linear model, while extensions are
discussed in Section~\ref{sec.GLM}:
%
%e2.1 #&#
\begin{equation}
\label{mod.lin} Y = \bx\beta^0 + \eps,
\end{equation}
with $n \times p$ fixed or random design matrix $\bx$, $n \times
1$ response and error
vectors $Y$ and $ \eps$, respectively. The errors are assumed to be
independent of $\bx$ (for random design) with i.i.d. entries having
$\EE[\eps_i] = 0$. We allow for high-dimensional settings where $p \gg
n$.
In further development, the active set or the set of relevant variables
\[
S_0 = \bigl\{j;\beta^0_j \neq0, j=1,
\ldots,p\bigr\},
\]
as well as its cardinality $s_0 = |S_0|$, are important quantities. The main
goals of this section are the construction of confidence intervals and
$p$-values for individual regression parameters $\beta^0_j (j=1,\ldots
,p)$ and corresponding multiple testing adjustment. The former is a highly
nonstandard problem in high-dimensional settings, while for the latter we
can use standard well-known techniques. When considering both
goals simultaneously, though, one can develop more powerful multiple testing
adjustments.
The Lasso (\cite{tibs96}) is among the most popular procedures for
estimating the unknown parameter $\beta^0$ in a high-dimensional linear
model. It exhibits desirable or sometimes even optimal properties for point
estimation such as prediction of $\bx\beta^0$ or of a new response
$Y_{\mathrm{new}}$, estimation in terms of $\|\hat{\beta} - \beta^0\|_q$
for $q = 1,2$, and variable selection or screening; see, for example,
the book of \citet{pbvdg11}. For assigning uncertainties in terms of
confidence intervals or hypothesis testing, however, the plain Lasso seems
inappropriate. It is very difficult to characterize the distribution of the
estimator in the high-dimensional setting; \citet{knfu00} derive asymptotic
results for fixed dimension as sample size $n \to\infty$ and already for
such simple situations, the asymptotic distribution of the Lasso has point
mass at zero. This implies, because of noncontinuity of the distribution,
that standard bootstrapping and subsampling schemes are delicate to apply
and uniform convergence to the limit seems hard to achieve. The latter
means that the estimator is exposed to undesirable super-efficiency
problems, as illustrated in Section~\ref{subsec.comparlm}. All the problems
mentioned are
expected to apply not only for the Lasso but also for other sparse
estimators as
well.

In high-dimensional settings and for general fixed design $\bx$, the
regression parameter is not identifiable. However, when making some
restrictions on the design, one can ensure that the regression vector is
identifiable. The so-called compatibility condition on the design $\bx$
(\cite{vandeGeer:07a}) is a rather weak assumption (\cite{van2009conditions})
which guarantees identifiability and oracle (near) optimality results for
the Lasso. For the sake of completeness, the compatibility condition is
described in Appendix~\ref{subsec.appadd}.

When assuming the compatibility condition with constant $\phi_0^2$
($\phi_0^2$ is close to zero for rather ill-posed designs, and sufficiently
larger than zero for well-posed designs), the
Lasso has the following property: for Gaussian errors and if $\lambda
\asymp\sqrt{\log(p)/n}$, we have with high probability that
%
%e2.2 #&#
\begin{equation}
\label{lasso-ell1} \bigl\|\hat{\beta} - \beta^0\bigr\|_1 \le4
s_0 \lambda/\phi_0^2.
\end{equation}
Thus, if $s_0 \ll\sqrt{n/\log(p)}$ and $\phi_0^2 \ge M > 0$, we have
$\|\hat{\beta} - \beta^0\|_1 \to0$ and, hence, the parameter $\beta^0$ is
identifiable.

Another often used assumption, although not necessary by any means, is the
so-called beta-min assumption:
%
%e2.3 #&#
\begin{equation}
\label{beta-min} \min_{j \in S_0}\bigl |\beta^0_j\bigr|
\ge\beta_{\mathrm{min}},
\end{equation}
for some choice of constant $\beta_{\mathrm{min}} > 0$.
The result in (\ref{lasso-ell1}) immediately implies the screening
property: if
$\beta_{\mathrm{min}} > 4 s_0 \lambda/\phi_0^2$, then
%
%e2.4 #&#
\begin{equation}
\label{screening} \hat{S} = \{j; \hat{\beta}_j \neq0\} \supseteq
S_0.
\end{equation}
Thus, the screening property holds when assuming the compatibility and
beta-min condition. The power of the screening property is a massive
dimensionality reduction (in the original variables) because $|\hat{S}|
\le
\min(n,p)$; thus, if $p \gg n$, the selected set $\hat{S}$ is
much smaller than the full set of $p$ variables. Unfortunately, the
required conditions are overly restrictive and exact variable screening
seems rather unrealistic in practical applications (\cite{pbmand13}).

%s2.1 #&#
\subsection{Different Methods}\label{subsec.lm-methods}

We describe here three different methods for construction of statistical
hypothesis tests or confidence intervals. Alternative procedures are
presented in Sections~\ref{subsec.othermeth} and \ref{subsec.comparlm}.

%s2.1.1 #&#
\subsubsection{Multi sample-splitting}\label{subsec.multisample-split}

A generic way for deriving $p$-values in hypotheses testing is given by
splitting the sample with indices $\{1,\ldots,n\}$ into two equal halves
denoted by $I_1$ and $I_2$, that is,
$I_r \subset\{1,\ldots,n\}\ (r=1,2)$ with $|I_1| = \lfloor n/2 \rfloor$,
$|I_2| = n - \lfloor n/2 \rfloor$, $I_1 \cap I_2 = \varnothing$ and $I_1
\cup
I_2 = \{1,\ldots, n\}$. The idea is to use the first half $I_1$ for variable
selection and the second half $I_2$ with the reduced set of selected
variables (from $I_1$) for statistical inference in terms of $p$-values. Such
a sample-splitting procedure avoids the over-optimism to use the data
twice for selection and inference after selection (without taking the effect
of selection into account).

Consider a method for variable selection based on
the first half of the sample:
\[
\hat{S}(I_1) \subset\{1,\ldots,p\}.
\]
A prime example is the Lasso which selects all the variables whose
corresponding estimated regression coefficients are different from
zero. We then use the second half of the sample $I_2$ for constructing
$p$-values, based on the selected variables $\hat{S}(I_1)$. If the
cardinality $|\hat{S}(I_1)| \le n/2 \le|I_2|$, we can run
ordinary least squares estimation using the subsample $I_2$ and the
selected variables $\hat{S}(I_1)$, that is, we regress $Y_{I_2}$ on
$\bx_{I_2}^{(\hat{S}(I_1))}$ where the sub-indices denote the sample
half and
the super-index stands for the selected variables, respectively. Thereby,
we implicitly assume that
the matrix $\bx_{I_2}^{(\hat{S}(I_1))}$ has full rank $|\hat{S}(I_1)|$. Thus,
from such a procedure, we
obtain $p$-values $P_{t\mbox{-}\mathrm{test},j}$ for testing $H_{0,j}: \beta^0_j
= 0$, for $j \in\hat{S}(I_1)$, from the classical $t$-tests,
assuming Gaussian errors or relying on asymptotic justification by the
central limit theorem. To be more precise, we define (raw) $p$-values
\begin{eqnarray*}
P_{\mathrm{raw},j} = \cases{ P_{t\mbox{-}\mathrm{test},j} \mbox{ based on $Y_{I_2},
\bx_{I_2}^{(\hat{S}(I_1))}$},\vspace*{2pt}\cr
 \quad \hspace*{10pt}\mbox{if} j \in\hat {S}(I_1),
\vspace*{2pt}
\cr
1, \quad \mbox{if} j \notin\hat{S}(I_1).}
\end{eqnarray*}
An interesting feature of such a sample-splitting procedure is the
adjustment for multiple testing. For example, if we wish to control the
familywise error rate over all considered hypotheses $H_{0,j}
(j=1,\ldots
,p)$, a naive approach would employ a Bonferroni--Holm correction over the
$p$ tests. This is not necessary: we only need to control over the
considered $|\hat{S}(I_1)|$ tests in $I_2$. Therefore, a Bonferroni
corrected $p$-value for $H_{0,j}$ is given by
\[
P_{\mathrm{corr},j} = \min\bigl(P_{\mathrm{raw},j} \cdot\bigl|\hat{S}(I_1)\bigr|,1
\bigr).
\]
In high-dimensional scenarios, $p \gg n > \lfloor n/2 \rfloor\geq
|\hat{S}(I_1)|$, where the latter inequality is an implicit assumption
which holds for the Lasso (under weak assumptions), and thus, the
correction factor employed here is rather
small.
Such corrected $p$-values control the familywise error rate in multiple
testing when assuming the screening property in (\ref{screening}) for the
selector $\hat{S} = \hat{S}(I_1)$ based on the first half $I_1$ only,
exactly as stated in Fact~\ref{th1} below. The reason is that the
screening property ensures that the reduced model is a correct model, and
hence the result is not surprising. In
practice, the screening property typically
does not hold exactly, but it is not a necessary condition for constructing
valid $p$-values (\cite{pbmand13}).

The idea about sample-splitting and subsequent statistical inference is
implicitly contained in \citet{WR08}. We summarize the whole procedure as
follows:

%\begin{center}
\emph{Single sample-splitting for multiple testing of $H_{0,j}$ among
$j=1,\ldots,p$}:
%\end{center}

%
\begin{longlist}[1.]
\item[1.] Split (partition) the sample $\{1,\ldots,n\} = I_1 \cup I_2$ with
$I_1 \cap I_2
= \varnothing$ and $|I_1| = \lfloor n/2 \rfloor$ and $|I_2| = n - \lfloor n/2
\rfloor$.
\item[2.] Using $I_1$ only, select the variables $\hat{S} \subseteq\{
1,\ldots
,p\}$. Assume or enforce that $|\hat{S}| \le|I_1| = \lfloor n/2 \rfloor
\le|I_2|$.
\item[3.] Denote the design matrix with the selected set of variables
by $\bx^{(\hat{S})}$. Based on $I_2$ with data
$(Y_{I_2},\bx_{I_2}^{(\hat{S})})$, compute $p$-values $P_{\mathrm
{raw,j}}$ for
$H_{0,j}$, for $j \in\hat{S}$, from classical least squares estimation
[i.e., $t$-test which can be used since $|\hat{S}(I_1)| \le|I_2|$]. For $j
\notin\hat{S}$, assign $P_{\mathrm{raw},j} = 1$.
\item[4.] Correct the $p$-values for multiple testing: consider
\[
P_{\mathrm{corr},j} = \min\bigl(P_j \cdot|\hat{S}|,1\bigr),
\]
which is an adjusted $p$-value for $H_{0,j}$ for controlling the familywise
error rate.
\end{longlist}
%
%\begin{algorithm}[h]
%\begin{algorithmic}[1]
%\STATE Split the sample $\{1,\ldots,n\} = I_1 \cup I_2$ with $I_1
%\cap I_2
%= \varnothing$ and $|I_1| = \lfloor n/2 \rfloor$ and $|I_2| = n -
%\lfloor n/2
%\rfloor$.
%\STATE Based on $I_1$, select the variables $\hat{S} \subseteq\{1,
%\ldots
%,p\}$. Assume (or ensure) that $|\hat{S}| \le|I_1| = \lfloor n/2
%\rfloor
%\le|I_2|$.
%\STATE Consider the reduced set of variables with design matrix
%$\bx^{(\hat{S})}$. Based on $I_2$ with data
%$(Y_{I_2},\bx_{I_2}^{(\hat{S})})$, compute p-values $P_{
%\mathrm{raw,j}}$ for
%$H_{0,j}$, for $j \in\hat{S}$, from classical least squares estimation
%(i.e. t-test which is reasonable since $|\hat{S}(I_1)| \le|I_2|$).
%For $j
%\notin\hat{S}$, assign $P_{\mathrm{raw},j} = 1$.
%\STATE Correct the p-values for multiple testing: consider
%\begin{eqnarray*}
%P_{\mathrm{corr},j} = \min(P_j \cdot|\hat{S}|,1)
%\end{eqnarray*}
%which is an adjusted p-value for $H_{0,j}$ for controlling the
%familywise
%error rate.
%\end{algorithmic}
%\caption{Single sample-splitting for multiple testing of $H_{0,j}$
%among
% $j=1,\ldots,p$: }\label{alg1}
%\end{algorithm}

%
%f1 #&#
\begin{figure}

\includegraphics{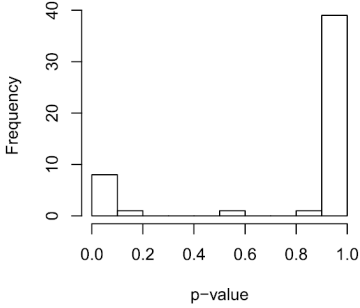}

\caption{Histogram of $p$-values $P_{\mathrm{corr},j}$ for a single
covariable, in the \texttt{riboflavin} data set, when doing 50
different (random) sample splits. The figure is taken from
B{\"{u}}hlmann, Kalisch and Meier (\citeyear{bumeka13}).}
\label{fig:pval_lottery}
\end{figure}

A major problem of the single sample-splitting method is its sensitivity
with respect to the choice of splitting the entire sample: sample
splits lead to
wildly different $p$-values. We call this undesirable phenomenon a $p$-value
lottery, and Figure~\ref{fig:pval_lottery} provides an illustration.
To overcome the ``$p$-value lottery,'' we can run the sample-splitting method
$B$ times, with $B$ large. Thus, we obtain a collection of $p$-values for the
$j$th hypothesis $H_{0,j}$:
\[
P_{\mathrm{corr},j}^{[1]},\ldots,P_{\mathrm{corr},j}^{[B]}\quad (j=1,
\ldots,p).
\]
The task is now to do an aggregation to a single $p$-value. Because of
dependence among $\{P_{\mathrm{corr},j}^{[b]}; b=1,\ldots,B\}$, because
all the different half samples are part of the same full sample, an
appropriate aggregation needs to be developed.
A simple solution is to use an empirical $\gamma$-quantile with $0 <
\gamma
< 1$:
\begin{eqnarray*}
&&Q_j(\gamma)\\
&&\quad = \min \bigl(\mbox{emp. $\gamma$-quantile}\bigl
\{P_{\mathrm{corr},j}^{[b]}/\gamma ; b=1,\ldots,B\bigr\},\\
&&\qquad 1 \bigr).
\end{eqnarray*}
For example, with $\gamma= 1/2$, this amounts to taking the sample median
$\{P_{\mathrm{corr},j}^{[b]}; b=1,\ldots,B\}$ and multiplying it with the
factor 2. A bit more sophisticated approach is to choose the best and
properly scaled $\gamma$-quantile in the
range $(\gamma_{\mathrm{min}},1)$ (e.g., $\gamma_{\mathrm{min}} = 0.05$),
leading to the aggregated $p$-value
%
%e2.5 #&#
\begin{eqnarray}
\label{aggreg} P_j = \min \Bigl(\bigl(1 - \log(\gamma_{\mathrm{min}})
\bigr) \inf_{\gamma\in
(\gamma_{\mathrm{min}},1)} Q_j(\gamma) \Bigr)
\nonumber
\\[-8pt]
\\[-8pt]
\eqntext{(j=1,
\ldots,p).}
\end{eqnarray}
Thereby, the factor $(1 - \log(\gamma_{\mathrm{min}}))$ is the price to be
paid for searching for the best $\gamma\in
(\gamma_{\mathrm{min}},1)$. This Multi sample-splitting procedure has been
proposed and analyzed in \citet{memepb09}, and we summarize it below. Before
doing so, we remark that
the aggregation of dependent $p$-values as described above is a general
principle as described in Appendix~\ref{subsec.appadd}.

%\begin{center}
\emph{Multi sample-splitting for multiple testing of $H_{0,j}$ among
$j=1,\ldots,p$}:
%\end{center}

\begin{longlist}[1.]
\item[1.] Apply the single sample-splitting procedure $B$ times,
leading to $p$-values $\{P_{\mathrm{corr},j}^{[b]}; b=1,\ldots
,B\}$. Typical choices are $B=50$ or $B=100$.
\item[2.] Aggregate these $p$-values as in (\ref{aggreg}), leading to
$P_{j}$ which are adjusted $p$-values for $H_{0,j} (j=1,\ldots,p)$,
controlling the familywise error rate.
\end{longlist}
%
%\begin{algorithm}[h]
%\begin{algorithmic}[1]
%\STATE Run the single sample-splitting Algorithm \ref{alg1} $B$ times
%leading to p-values $\{P_{\mathrm{corr},j}^{[b]}; b=1,\ldots,B\}$. A
%typical choice is $B=100$.
%\STATE Aggregate the p-values from Step 1 as in (\ref{aggreg}) leading
%to
%$P_{j}$ which are adjusted p-values for $H_{0,j} (j=1,\ldots,p)$,
%controlling the familywise error rate.
%\end{algorithmic}
%\caption{Multi sample-splitting for multiple testing of $H_{0,j}$ among
% $j=1,\ldots,p$}\label{alg2}
%\end{algorithm}
The Multi sample-splitting method enjoys the property that the resulting
$p$-values are approximately reproducible and not subject to a ``$p$-value lottery''
anymore, and it controls the
familywise error rate under the following assumptions:
\begin{longlist}[(A1)]
\item[(A1)] The screening\vspace*{1pt} property as in (\ref{screening}) for the
first half of
the sample: $\PP[\hat{S}(I_1) \supseteq S_0] \ge1 - \delta$ for some
$0 < \delta< 1$.
\item[(A2)] The reduced design matrix for the second half of the sample
satisfies
$\mathrm{rank}(\bx_{I_2}^{(\hat{S}(I_1))}) = |\hat{S}(I_1)|$.
\end{longlist}
%
%fa1 #&#
\begin{theo}[{[\citet{memepb09}]}]\label{th1}
Consider a linear model as in (\ref{mod.lin}) with fixed design $\bx$ and
Gaussian errors. Assume \textup{(A1)--(A2)}.
Then, for a significance level $0 < \alpha< 1$ and denoting by $B$ the
number of sample splits,
\[
\PP\biggl[\bigcup_{j \in S_0^c} I(P_j \le
\alpha)\biggr] \le\alpha+ B \delta,
\]
that is, the familywise error rate (FWER) is controlled up to the
additional (small) value $B \delta$.
\end{theo}

A proof is given in Meinshausen, Meier and  B{\"u}hlmann
(\citeyear{memepb09}). We note that the Multi
sample-splitting method can be used in conjunction with any reasonable,
sparse
variable screening method fulfilling (A1) for very small $\delta> 0$ and
(A2); and it does not necessarily rely on the Lasso for variable
screening. See also Section~\ref{subsec.othersparsemeth}.\vspace*{1pt} Assumption (A2) typically holds for the Lasso
satisfying $|\hat{S}(I_1)| \le|I_1| = \lfloor n/2 \rfloor\le|I_2| =
n -
\lfloor n/2 \rfloor$.

\emph{The screening property} (A1). The screening property (A1) with very
small $\delta> 0$ is not a
necessary condition for constructing valid $p$-values and can be replaced by
a zonal assumption requiring the following: there is a gap between large
and small regression coefficients and there are not too many small nonzero
regression coefficients (\cite{pbmand13}). Still, such a zonal assumption
makes a requirement about the unknown $\beta^0$ and the absolute values of
its components: but this is the essence of the question in hypothesis
testing to infer whether coefficients are sufficiently different from zero,
and one would like to do such a test without an assumption on the true
values.

The Lasso satisfies (A1) with $\delta\to0$ when
assuming the compatibility condition (\ref{compat}) on the design $\bx$,
the sparsity assumption $s_0 = o(\sqrt{n/\log(p)})$ [or $s_0 =
o(n/\log(p))$ when requiring a restricted eigenvalue assumption] and a
beta-min
condition (\ref{beta-min}), as shown in
(\ref{screening}). Other procedures also exhibit the screening
property such as the adaptive Lasso (\cite{zou06}), analyzed in detail in
\citet{geer11}, or methods with concave regularization penalty such as
SCAD (\cite{fan2001variable}) or MC$+$ (\cite{zhang2010}). As criticized
above, the required beta-min assumption should be avoided when
constructing a hypothesis test about the unknown components of $\beta^0$.

Fact~\ref{th1} has a corresponding asymptotic formulation
where the dimension $p = p_n$ and the model depends on sample size $n$: if
(A1) is replaced by $\lim_{n \to\infty} \PP[\hat{S}(I_{1;n}) \supseteq
S_{0;n}] \to1$ and for a fixed number $B$, $\limsup_{n \to\infty}
\PP[\bigcup_{j \in S_0^c} I(P_j \le\alpha)] \le\alpha$.
%Again, the Lasso
%satisfies the asymptotic
%screening property when assuming asymptotic versions of the
%compatibility
%and beta-min assumptions.
In such an asymptotic setting, the Gaussian
assumption in Fact~\ref{th1} can be relaxed by invoking the central
limit theorem (for the low-dimensional part).

The Multi sample-splitting method is very generic: it can be used for many
other models, and its basic assumptions are an approximate screening property
(\ref{screening}) and that the cardinality $|\hat{S}(I_1)| < |I_2|$ so that
we only have to deal with a fairly low-dimensional inference problem. See,
for example, Section~\ref{sec.GLM} for GLMs. An extension for testing group
hypotheses of the form $H_{0,G}: \beta_j = 0$ for all $j
\in G$ is indicated in Section~\ref{subsec.assfree}.

Confidence intervals can be constructed based on the duality with the
$p$-values from equation (\ref{aggreg}). A procedure is described in detail
in Appendix~\ref{subsec.appmssplitci}.
The idea to invert the $p$-value method is to apply a bisection method having
a point in and a point outside of the confidence interval. To verify if a
point is inside the
\emph{aggregated} confidence interval, one looks at the fraction of
confidence intervals from the splits which cover the point.
%Each
%confidence interval from a different split will have a different
%confidence level, depending on the ordering of the p-values of all the
%splits.

%s2.1.2 #&#
\subsubsection{Regularized projection: De-sparsifying the
Lasso}\label{subsec.desparslasso}

We describe here a method, first introduced by \citet{zhangzhang11}, which
does not require an assumption about $\beta^0$ except for sparsity.

It is instructive to give a motivation starting with the low-dimensional
setting where $p < n$ and $\mathrm{rank}(\bx) = p$. The $j$th component of
the ordinary least squares estimator $\hat{\beta}_{\mathrm{OLS};j}$ can
be obtained
as follows. Do an OLS regression of $\bx^{(j)}$ versus all other variables
$\bx^{(-j)}$ and denote the corresponding residuals by $Z^{(j)}$. Then
%
%e2.6 #&#
\begin{equation}
\label{proj-ols} \hat{\beta}_{\mathrm{OLS};j} = Y^T Z^{(j)}/
\bigl(\bx^{(j)}\bigr)^T Z^{(j)}
\end{equation}
can be obtained by a linear projection.
In a high-dimensional setting, the residuals $Z^{(j)}$ would be equal to
zero and the projection is ill-posed.

For the high-dimensional case with $p > n$, the idea is to pursue a
regularized projection. Instead of ordinary least squares regression, we
use a Lasso regression of $\bx^{(j)}$ versus $\bx^{(-j)}$ with
corresponding residual vector $Z^{(j)}$: such a penalized regression
involves a
regularization parameter $\lambda_j$ for the Lasso, and hence $Z^{(j)} =
Z^{(j)}(\lambda_j)$. As in (\ref{proj-ols}), we immediately obtain (for any
vector $Z^{(j)}$)
%
%e2.7 #&#
\begin{eqnarray}
\label{proj-lasso} \qquad \frac{Y^T Z^{(j)}}{(\bx^{(j)})^T Z^{(j)}} &=& \beta^0_j + \sum
_{k \neq
j} P_{jk} \beta^0_k +
\frac{\eps^T Z^{(j)}}{(\bx^{(j)})^T Z^{(j)}},
\nonumber
\\[-8pt]
\\[-8pt]
\nonumber
P_{jk}&=& \bigl(\bx^{(k)}\bigr)^T
Z^{(j)}/\bigl(\bx^{(j)}\bigr)^T Z^{(j)}.
\end{eqnarray}
We note that in the low-dimensional case with $Z^{(j)}$ being the residuals
from ordinary least squares, due to orthogonality, $P_{jk} = 0$. When using
the Lasso-residuals for $Z^{(j)}$, we do not have exact orthogonality
and a
bias arises. Thus, we make a bias correction by plugging in the Lasso
estimator $\hat{\beta}$ (of the regression $Y$ versus $\bx$): the
bias-corrected estimator is
%
%e2.8 #&#
\begin{equation}
\label{despars-lasso} \hat{b}_j = \frac{Y^T Z^{(j)}}{(\bx^{(j)})^T Z^{(j)}} - \sum
_{k \neq j} P_{jk} \hat{\beta}_k.
\end{equation}
Using (\ref{proj-lasso}), we obtain
\begin{eqnarray*}
\sqrt{n}\bigl(\hat{b}_j - \beta^0_j\bigr)
&= &\frac{n^{-1/2} \eps^T Z^{(j)}}{n^{-1}
(\bx^{(j)})^T Z^{(j)}}\\
&&{} + \sum_{k \neq j} \sqrt{n}
P_{jk}\bigl(\beta_k^0 - \hat{\beta}_k
\bigr).
\end{eqnarray*}
The first term on the right-hand side has a Gaussian
distribution, when assuming Gaussian errors; otherwise, it has an
asymptotic Gaussian distribution assuming that $\EE|\eps_i|^{2 + \kappa}
< \infty$ for $\kappa> 0$ (which suffices for the Lyapunov CLT). We will
argue in Appendix~\ref{subsec.appadd} that the second term is negligible
under the following assumptions:
\begin{longlist}[(B1)]
\item[(B1)] The design matrix $\bx$ has compatibility constant bounded away
from zero, and the sparsity is $s_0 = o(\sqrt{n}/\log(p))$.
\item[(B2)] The rows of $\bx$ are fixed realizations of i.i.d. random
vectors $\sim{\cal N}_p(0,\Sigma)$, and the minimal eigenvalue of
$\Sigma$ is bounded away from zero.
\item[(B3)] The inverse $\Sigma^{-1}$ is row-sparse with $s_j =
\sum_{k \neq j} I((\Sigma^{-1})_{jk} \neq0) =
o(n/\log(p))$.
\end{longlist}
%
%fa2 #&#
\begin{theo}[(\cite{zhangzhang11};
van~de Geer et\break al., \citeyear{vdgetal13})]\label{th2}
Consider a linear model as in (\ref{mod.lin}) with fixed design and
Gaussian errors. Assume \textup{(B1)}, \textup{(B2)} and \textup{(B3)} (or an $\ell_1$-sparsity
assumption on the rows of $\Sigma^{-1}$).
%($\max_j \|\gamma^0_j\|_1 =
% O(\sqrt{n/\log(p)})$).}
Then
\begin{eqnarray*}
\sqrt{n} \sigma_{\eps}^{-1} \bigl(\hat{b} -
\beta^0\bigr) &=& W + \Delta,\quad W \sim{\cal N}_p(0,\Omega),
\\
\Omega_{jk} &=&
\frac{n(Z^{(j)})^T Z^{(k)}}{[(\bx^{(j)})^T Z^{(j)}][(X^{(k)})^T
Z^{(k)}]},
\\
\|\Delta\|_{\infty} &=& o_P(1).
\end{eqnarray*}
[We note that this statement holds with probability tending to one, with
respect to the variables $\bx\sim{\cal N}_P(0,\Sigma)$ as assumed in
\textup{(B2)}].
\end{theo}

The asymptotic implications of Fact~\ref{th2} are as follows:
\[
\sigma_{\eps}^{-1} \Omega_{jj}^{-1/2}
\sqrt{n} \bigl(\hat{b}_j - \beta^0_j\bigr)
\Rightarrow{\cal N}(0,1),
\]
from which we can immediately construct a confidence interval or hypothesis
test by plugging in an estimate $\hat{\sigma}_{\eps}$ as briefly discussed
in Section~\ref{subsec.addissues}. From a theoretical perspective, it is more
elegant to use the square root Lasso (\cite{belloni2011square}) for the
construction of $Z^{(j)}$; then one can drop (B3) [or the
$\ell_1$-sparsity version
of (B3)] (\cite{vdg14}). In fact, all that we then need is formula
(\ref{ell1bound})
\[
\bigl\|\hat{\beta} - \beta^0\bigr\|_1 = o_P\bigl(1/
\sqrt{\log(p)}\bigr).
\]
From a practical perspective, it seems to make essentially no difference
whether\vspace*{1pt} one takes the square root or plain Lasso for the construction of
the $Z^{(j)}$'s.

More general than the statements in Fact~\ref{th2}, the following
holds assuming (B1)--(B3) (\cite{vdgetal13}): the asymptotic variance
$\sigma_{\eps}^2 \Omega_{jj}$
reaches the Cram\'{e}r--Rao lower bound, which equals $\sigma_{\eps}^2
(\Sigma^{-1})_{jj}$ [which is bounded away from zero, due to (B2)], and the
estimator $\hat{b}_j$ is efficient in the sense
of semiparametric inference. Furthermore, the convergence in Fact~\ref{th2} is uniform over the subset of the parameter space where the
number of nonzero coefficients $\|\beta^0\|_0$ is small and, therefore, we
obtain \emph{honest} confidence intervals and tests. In particular,
both of
these results say that all the complications in post-model
selection do not arise (\cite{leebpoetsch03}), and yet $\hat{b}_j$ is
optimal for construction of confidence intervals of a single coefficient
$\beta^0_j$.

From a practical perspective, we need to choose the regularization
parameters $\lambda$ (for the Lasso regression of $Y$ versus $\bx$) and
$\lambda_j$ [for the nodewise Lasso regressions (\cite{mebu06}) of $\bx^{(j)}$
versus all other variables $\bx^{(-j)}$]. Regarding the former, we
advocate a choice using cross-validation; for the latter, we favor a
proposal for a smaller
$\lambda_j$ than the one from CV, and the details are described in Appendix~\ref{subsec.appadd}.

Furthermore, for a group $G \subseteq\{1,\ldots
,p\}$, we can test
a group hypothesis $H_{0,G}: \beta^0_j = 0$ for all $j \in G$ by
considering the test-statistic
\[
\max_{j \in G} \sigma_{\eps}^{-1}
\Omega_{jj}^{-1/2} \sqrt{n} |\hat{b}_j| \Rightarrow
\max_{j \in G} \Omega_{jj}^{-1/2}
|W_j|,
\]
where the limit on the right-hand side occurs if the null-hypothesis
$H_{0,G}$ holds true.
The distribution of $\max_{j \in G} |\Omega_{jj}^{-1/2} W_j|$ can be easily
simulated from dependent Gaussian random variables. We also remark that
sum-type statistics for large groups
cannot be easily treated because $\sum_{j \in G} |\Delta_j|$ might get out
of control.

%s2.1.3 #&#
\subsubsection{Ridge projection and bias correction}\label{subsec.ridge-proj}

Related to the desparsified Lasso estimator $\hat{b}$ in
(\ref{despars-lasso}) is an approach based on Ridge estimation. We sketch
here the main properties and refer to \citet{pb13} for a detailed
treatment.

Consider
\[
\hat{\beta}_{\mathrm{Ridge}} = \bigl(n^{-1} \bx^T \bx+
\lambda I\bigr)^{-1} n^{-1} \bx^T Y.
\]
A major source of bias occurring in Ridge estimation when $p > n$ comes
from the fact that the Ridge estimator is estimating a projected parameter
\[
\theta^0 = P_{R} \beta^0,\quad P_{R} =
\bx^T \bigl(\bx\bx^T\bigr)^{-}\bx,
\]
where $(\bx\bx^T)^{-}$ denotes a generalized inverse of $\bx\bx^T$. The
minor bias for $\theta^0$ then satisfies
\begin{eqnarray*}
\max_j\bigl|\EE[\hat{\beta}_{\mathrm{Ridge};j}] -
\theta^0_j\bigr| \le\lambda\bigl\| \theta^0
\bigr\|_2 \lambda_{\mathrm{min} \neq0}(\hat{\Sigma})^{-1},
\end{eqnarray*}
where $\lambda_{\mathrm{min} \neq0}(\hat{\Sigma})$ denotes the minimal
nonzero eigenvalue of $\hat{\Sigma}$ (\cite{shadeng11}). The quantity can
be made small by choosing $\lambda$ small. Therefore, for
$\lambda\searrow0^+$ and assuming Gaussian errors, we have that
%
%e2.9 #&#
\begin{equation}
\label{Ridge-distr}\quad \sigma_{\eps}^{-1} \bigl(\hat{
\beta}_{\mathrm{Ridge}} - \theta^0\bigr) \approx W,\quad W \sim{\cal
N}_p(0, \Omega_R),
\end{equation}
where $\Omega_R = (\hat{\Sigma} + \lambda)^{-1} \hat{\Sigma} (\hat
{\Sigma} +
\lambda)^{-1}/n$. Since
\[
\frac{\theta^0}{P_{R;jj}} = \beta^0_j + \sum
_{k \neq j} \frac{P_{R;jk}}{P_{R;jj}} \beta^0_k,
\]
the major bias for $\beta^0_j$ can be estimated and corrected with
\[
\sum_{k \neq j} \frac{P_{R;jk}}{P_{R;jj}} \hat{
\beta}_k,
\]
where $\hat{\beta}$ is the ordinary Lasso. Thus, we construct a
bias-corrected Ridge estimator, which addresses the potentially substantial
difference between $\theta^0$ and the target $\beta^0$:
%
%e2.10 #&#
\begin{eqnarray}
\label{corr-Ridge} \hat{b}_{R;j} = \frac{\hat{\beta}_{\mathrm{Ridge};j}}{P_{R;jj}} - \sum
_{k \neq
j} \frac{P_{R;jk}}{P_{R;jj}} \hat{\beta}_k,
\nonumber
\\[-8pt]
\\[-8pt]
 \eqntext{j=1,
\ldots,p.}
\end{eqnarray}
Based on (\ref{Ridge-distr}), we derive in Appendix~\ref{subsec.appadd} that
%
%e2.11 #&#
\begin{eqnarray}
\label{Ridge-repr}&&\sigma_{\eps}^{-1} \Omega_{R;jj}^{-1/2}
\bigl(\hat{b}_{R;j} - \beta^0_j\bigr)\nonumber\\
&&\quad \approx
\Omega_{R;jj}^{-1/2} W_j / P_{R;jj}\nonumber \\
&&\qquad{}+
\sigma_{\eps}^{-1} \Omega_{R;jj}^{-1/2}
\Delta_{R;j},\quad
W \sim{\cal N}_p(0, \Omega_R),
\\
&&|\Delta_{R;j}| \le\Delta_{R\mathrm{bound};j} \nonumber\\
&&\hspace*{22pt}\quad:= \max_{k \neq
j}
\biggl\llvert \frac{P_{R;jk}}{P_{R;jj}}\biggr\rrvert \bigl(\log(p)/n\bigr)^{1/2 - \xi},
\nonumber\end{eqnarray}
with the typical choice $\xi= 0.05$. Sufficient conditions for deriving
(\ref{Ridge-repr}) are assumption (B1) and that the sparsity satisfies $s_0
=O((n/\log(p))^{\xi})$ for $\xi$ as above.

Unlike as in Fact~\ref{th2}, the term $\Delta_{R;j}$ is typically not
negligible and we correct the Gaussian part in (\ref{Ridge-repr}) by the
upper bound $\Delta_{R\mathrm{bound};j}$. For example,
for testing $H_{0,j}: \beta^0_j = 0$ we use the upper bound for the $p$-value
\begin{eqnarray*}
2\bigl(1 - \Phi\bigl(\sigma_{\eps}^{-1}\Omega_{R;jj}^{-1/2}
|P_{R;jj}|\bigl(|\hat {b}_{R;j}| - \Delta_{R\mathrm{bound};j}\bigr)_+\bigr)
\bigr).
\end{eqnarray*}
Similarly, for two-sided confidence intervals with coverage $1-\alpha$
we use
\begin{eqnarray*}
& &[\hat{b}_{R;j} -c_j,\hat{b}_{R;j} +
c_j],
\\
& &c_j = \Delta_{R\mathrm{bound};j} + \sigma_{\eps} \Omega
_{R;jj}^{1/2}/|P_{R;jj}| \Phi^{-1}(1-
\alpha/2).
\end{eqnarray*}

For testing a group hypothesis for $G \subseteq\{1,\ldots,p\}$,
$H_{0,G}: \beta^0_j = 0$ for all $j \in G$, we can proceed similarly
as at
the end of Section~\ref{subsec.desparslasso}: under the null-hypotheses
$H_{0,G}$, the statistic $\sigma_{\eps}^{-1}
\max_{j \in G} \Omega_{R;jj}^{-1/2} |\hat{b}_{R;j}|$ has a distribution
which is approximately stochastically upper
bounded by
\begin{eqnarray*}
\max_{j \in G} \bigl(\Omega_{R;jj}^{-1/2}
|W_j| / |P_{R;jj}| + \sigma_{\eps}^{-1}
\Omega_{R;jj}^{-1/2} |\Delta_{R;j}|\bigr);
\end{eqnarray*}
see also (\ref{Ridge-repr}).
When invoking an upper bound for $\Delta_{R\mathrm{bound};j} \ge
|\Delta_{R;j}|$ as in (\ref{Ridge-repr}), we can easily simulate this
distribution from dependent Gaussian random variables, which in turn
can be
used to construct a $p$-value; we refer for further details to \citet{pb13}.

%s2.1.4 #&#
\subsubsection{Additional issues: Estimation of the error variance and
multiple testing correction}\label{subsec.addissues}

Unlike the Multi sample-splitting procedure in Section~\ref{subsec.multisample-split}, the desparsified Lasso and Ridge
projection method outlined in
Sections~\ref{subsec.desparslasso}--\ref{subsec.ridge-proj} require to
plug-in an estimate of $\sigma_{\eps}$ and to adjust for multiple
testing. The scaled Lasso (\cite{sunzhang11})
leads to a consistent estimate of the error variance: it is a fully automatic
method which does not need any specification of a tuning parameter. In
\citet{reidtibsh13}, an empirical comparison of various
estimators suggests that the estimator based on a residual sum of
squares of
a cross-validated Lasso solution often yields good finite-sample
performance.

Regarding the adjustment when doing many tests for individual regression
parameters or groups thereof, one can use any valid standard
method to correct the $p$-values from the desparsified Lasso or Ridge
projection method. The prime examples are the Bonferroni--Holm procedure for
controlling the familywise error rate and the method from \citet{benyek01}
for controlling the false discovery rate. An approach for
familywise error control which explicitly takes the dependence among the
multiple hypotheses is proposed in \citet{pb13}, based on simulations for
dependent Gaussian random variables.

%s2.1.5 #&#
\subsubsection{Conceptual differences between the methods}

We briefly outline here conceptual differences while Section~\ref{subsec.comparlm} presents empirical results.

The Multi sample-splitting
method is very generic and in the spirit of Breiman's appeal for stability
(\citeauthor{brei96}, \citeyear{brei96,brei96b}), it enjoys some kind of stability due to multiple
sample splits and aggregation; see also the discussion in Sections~\ref{subsec.othersparsemeth} and \ref{subsec.mainass}. The disadvantage
is that, in the worst
case, the method needs a beta-min or a weaker zonal assumption on the
underlying regression parameters: this is somewhat unpleasant since a
significance test should
\emph{find out} whether a regression coefficient is sufficiently large or
not.

Both the desparsified Lasso and Ridge projection procedures do not make
any assumption on the underlying regression coefficient except
sparsity. The
former is most powerful and asymptotically optimal if the design were
generated from a population distribution whose inverse covariance
matrix is
sparse. Furthermore, the convergence is uniform over all sparse regression
vectors and, hence, the method yields honest confidence regions or tests.
The Ridge projection method does not require any assumption on the
fixed design but does not reach the asymptotic Cram\'{e}r--Rao efficiency
bound. The construction with the additional correction term in
(\ref{delta-bound}) leads to reliable type I error control at the cost of
power.

In terms of computation, the Multi sample-splitting and Ridge projection
method are substantially less demanding than the desparsified Lasso.

%s2.1.6 #&#
\subsubsection{Other sparse methods than the
Lasso}\label{subsec.othersparsemeth}

All the methods described above are used ``in default mode'' in conjunction
with the Lasso (see also Section~\ref{subsec.hdilin}). This is not
necessary, and other estimators can be used.

For the Multi sample-splitting procedure, assumptions (A1) with $\delta
\to
0$ and (A2) are
sufficient for asymptotic correctness; see Fact~\ref{th1}. These
assumptions hold for many reasonable sparse estimators when requiring a
beta-min assumption and some sort of identifiability condition such as the
restricted eigenvalue or the compatibility condition on the design matrix~$\bx$; see also the discussion after Fact~\ref{th1}. It is unclear whether
one could gain substantially by using a different screening method than the
Lasso. In fact, the Lasso has been empirically found to perform rather well
for screening in comparison to the elastic net
(\cite{zou2005regularization}), marginal correlation screening
(\cite{fanlv07}) or thresholded Ridge regression; see \citet{pbmand13}.

For the desparsified Lasso, the error of
the estimated bias correction can be controlled by using a bound for
$\|\hat{\beta} - \beta^0\|_1$. If we require (B2) and (B3) [or an $\ell_1$
sparsity assumption instead of (B3)], the estimation error in the bias
correction, based on an estimator
$\hat{\beta}$ in (\ref{despars-lasso}), is asymptotically negligible if
%
%e2.12 #&#
\begin{equation}
\label{ell1bound} \bigl\|\hat{\beta} - \beta^0\bigr\|_1 =
o_P\bigl(1/\sqrt{\log(p)}\bigr).
\end{equation}
This bound is implied by (B1) and (B2) for the Lasso, but other estimators
exhibit this bound as well, as mentioned below. When using such another
estimator, the wording ``desparsified Lasso'' does not make sense
anymore. Furthermore, when using the
square root Lasso for the construction of $Z^{(j)}$, we only need
(\ref{ell1bound}) to obtain asymptotic normality with the $\sqrt{n}$
convergence rate (\cite{vdg14}).

For the Ridge projection method, a bound for $\|\hat{\beta} - \beta^0\|_1$
is again the only assumption such that the procedure is asymptotically
valid. Thus, for the corresponding bias correction, other methods than the
Lasso can be used.

We briefly mention a few other methods for which we have reasons that (A1)
with very small $\delta> 0$ and (A2), or the bound in (\ref{ell1bound})
hold: the adaptive Lasso
(\cite{zou06}) analyzed in greater detail in \citet{geer11}, the MC$+$
procedure with its high-dimensional mathematical analysis
(\cite{zhang2010}), or methods with concave regularization penalty such as
SCAD (\cite{fan2001variable}) analyzed in broader generality and detail in
\citet{fan2014}. If the assumptions (A1) with small $\delta> 0$ and (A2)
fail for the Multi sample-splitting method, the multiple sample
splitting still allows to check the
stability of the $p$-values $P_{\mathrm{corr},j}^{[b]}$ across $b$ (i.e.,
across sample splits). If the variable screening is unstable, many of the
$P_{\mathrm{corr},j}^{[b]}$ (across $b$) will be equal to 1, therefore, the
aggregation has a tendency to produce small $p$-values if most of them, each
from a sample split, are stable and small. See also
\citet{manbu13}, Section~5. In connection with the desparsified method, a
failure of the single sufficient condition in (\ref{ell1bound}), when
using, for example, the square root Lasso for construction of the
$Z^{(j)}$'s, might result
in a too large bias. In absence of resampling or Multi sample
splitting, it
seems difficult to diagnose such a failure (of the desparsified or Ridge
projection method) with real data.

%s2.2 #&#
\subsection{\texttt{hdi} for Linear Models}\label{subsec.hdilin}
In the \textsf{R}-package \texttt{hdi}, available on R-Forge (\cite{hdipackage}), we provide implementations for the
Multi sample-splitting, the Ridge projection and the
desparsified Lasso method.

Using the \textsf{R} functions is straightforward:

\begin{verbatim}
> outMssplit
  <- multi.split(x = x, y = y)
> outRidge
  <- ridge.proj(x = x, y = y)
> outLasso
  <- lasso.proj(x = x, y = y)
\end{verbatim}

For users that are very familiar with the procedures, we provide flexible
options. For example, we can easily use an alternative model selection
or another
``classical'' fitting procedure using the arguments \texttt{model.selector}
and \texttt{classical.fit} in \texttt{multi.split}. The default options
should be satisfactory for standard usage.

All procedures return $p$-values and confidence intervals. The Ridge and
desparsified Lasso methods return both single testing $p$-values as well as
multiple testing corrected $p$-values, unlike the Multi sample-splitting
procedure which only returns multiple testing corrected $p$-values. The
confidence intervals are for individual parameters only (corresponding to
single hypothesis testing).

The single testing $p$-values and the multiple testing corrected
$p$-values are extracted from the fit as follows:

\begin{verbatim}
> outRidge$pval
> outRidge$pval.corr
\end{verbatim}

By default, we correct for controlling the familywise error rate for
the $p$-values \texttt{pval.corr}.

Confidence intervals are acquired through the usual \texttt{confint}
interface. Below we extract the 95 \% confidence intervals for those
$p$-values that are smaller than \texttt{0.05}:

\begin{verbatim}
> confint(outMssplit,
  parm = which(outMssplit$pval.corr
   <= 0.05),
   level = 0.95)
\end{verbatim}

Due to the fact that the desparsified Lasso method is quite computationally
intensive, we provide the option to parallelize the method on a
user-specified number of cores.

We refer to the manual of the package for more detailed information.

%s2.3 #&#
\subsection{Other Methods}\label{subsec.othermeth}

Recently, other procedures have been suggested for construction of
$p$-values and confidence intervals.

Residual-type bootstrap approaches are proposed and analyzed in
\citet{chatter13} and \citet{liuyu13}. A problem with these approaches
is the nonuniform convergence to a limiting distribution and exposure to
the super-efficiency phenomenon, that is, if the true parameter equals
zero, a confidence region might be the singleton $\{0\}$ (due to a finite
amount of bootstrap resampling), while for nonzero true parameter values,
the coverage might be very poor or a big length of the confidence interval.

The covariance test (\cite{covtest14}) is another proposal which
relies on the solution path of the Lasso and provides $p$-values for
conditional tests that all relevant variables enter the Lasso solution path
first. It is related to post-selection inference, mentioned in Section~\ref{subsec.postsel}.

In \citet{jamo13b}, a procedure was proposed that is very similar to the
one described in Section~\ref{subsec.desparslasso}, with the only
difference being that Z is picked
as the solution of a convex program rather than using the Lasso. The
method is aiming to relax the sparsity assumption (B3) for the design.

A conservative \emph{Group-bound} method which needs no regularity
assumption for the
design, for example, no compatibility assumption (\ref{compat}), has
been proposed
by \citet{meins13}. The method has the capacity to
automatically determine whether a regression coefficient is
identifiable or
not, and this makes the procedure very robust against ill-posed
designs. The
main motivation of the method is in terms of testing groups of correlated
variables, and we discuss it in more detail in Section~\ref{subsec.assfree}.

While all the methods mentioned above are considered in a comparative
simulation study in Section~\ref{subsec.comparlm}, we mention here some
others. The idea of estimating a low-dimensional component of a
high-dimensional parameter is also worked out in
\citet{belloni2012sparse}, \citet{beletal13}, bearing connections to the
approach of
desparsifying the Lasso. Based on stability selection
(\cite{mebu10}), \citet{shah13} propose a version which leads to $p$-values
for testing
individual regression parameters. Furthermore, there are new and
interesting proposals for controlling the false discovery rate, in a
``direct way'' (\citeauthor{bogdan13} \citeyear{bogdan13,bogdan14}; \cite{foygcand14}).

%s2.4 #&#
\subsection{Main Assumptions and Violations}\label{subsec.mainass}

We discuss here some of the main assumptions, potential violations and
some corresponding implications calling for caution when aiming for
confirmatory conclusions.

%pa2.4.subsubsection.1 #&#
\textit{Linear model assumption}. The first one is that the linear (or
some other) model is
correct. This might be rather unrealistic and, thus, it is important to
interpret the output of software or a certain method. Consider a nonlinear
regression model
\begin{eqnarray*}
&&\mbox{random design}:\quad Y_0 = f^0(X_0) +
\eta_0,
\\
&&\mbox{fixed design}:\quad Y = f^0(\bx) + \eta,
\end{eqnarray*}
where, with some slight abuse of notation, $f^0(\bx) = f^0(\bx
_1),\ldots,
(f^0(\bx_n))^T$. We assume for the random design model, $\eta_0$ is
independent from $X_0$, $\EE[\eta_0] = 0$, $\EE[f^0(X_0)] = 0$, $\EE
[X_0] =
0$, and the data are $n$
i.i.d. realizations of $(X_0,Y_0)$; for the fixed design model, the $n
\times1$ random vector $\eta$ has i.i.d. components with
$\EE[\eta_i]=0$. For the random design model, we consider
%
%e2.13 #&#
\begin{eqnarray}
\label{betaproj} Y_0 &=& \bigl(\beta^0\bigr)^T
X_0 + \eps_0,\nonumber \\
\eps_0 &=& f^0(X_0)
- \bigl(\beta^0\bigr)^T X_0 +
\eta_0,
\\
\beta^0 &=&\argmin_{\beta} \EE\bigl[\bigl(f^0(X_0)
- \beta^T X_0\bigr)^2\bigr]\nonumber
\end{eqnarray}
[where the latter is
unique if $\Cov(X_0)$ is positive definite]. We note that $\EE[\eps_0|X_0]
\neq0$ while $\EE[\eps_0] = 0$ and, therefore, the inference should be
\emph{unconditional} on $\bx$ and is to be interpreted for the projected
parameter $\beta^0$ in (\ref{betaproj}). Furthermore, for correct asymptotic
inference of the projected parameter $\beta^0$, a modified estimator for
the asymptotic variance of the estimator is needed; and then both the
Multi sample-splitting and the desparsified Lasso are asymptotically
correct (assuming similar conditions as if the model were correct). The
Multi sample-splitting method is well suited for the random design case
because the sample splitting (resampling type) is coping well with
i.i.d. data. This is in contrast to fixed design, where the data is not
i.i.d. and the Multi sample-splitting method for a misspecified linear
model is typically not working anymore. The details are given in
\citet{pbvdg15}.

For a fixed design model with $\mathrm{rank}(\bx) = n$, we can always write
\[
Y = \bx\beta^0 + \eps,\quad \eps= \eta
\]
for many solutions $\beta^0$. For ensuring that the inference is valid, one
should consider a sparse $\beta^0$, for example, the basis pursuit
solution from
compressed sensing (\cite{candes2006near}) as one among many
solutions. Thus, inference should be
interpreted for a \emph{sparse} solution $\beta^0$, in the sense that a
confidence interval for the $j$th component would cover this $j$th
component of all sufficiently sparse solutions $\beta^0$. For the
high-dimensional fixed design case,
there is no misspecification with respect to linearity of the model;
misspecification might happen, though, if there is no solution $\beta
^0$ which
fulfills a required sparsity condition. The details are given again in
\citet{pbvdg15}.

The assumption about constant error variance might not hold. We note that
in the random design case of a nonlinear model as above, the error in
(\ref{betaproj}) has nonconstant variance when conditioning on $\bx$, but,
unconditionally, the noise is homoscedastic. Thus, as outlined, the
inference for a random design linear model is asymptotically valid
(unconditional on $\bx$) even though the conditional error distribution
given $\bx$ has nonconstant variance.

%pa2.4.subsubsection.2 #&#
\textit{Compatibility or incoherence-type assumption}.
The methods in Section~\ref{subsec.lm-methods} require an identifiability
assumption such as the compatibility condition on the design matrix $\bx$
described in (\ref{compat}). The procedure in Section~\ref{subsec.assfree}
does not require such an assumption: if a component of the regression
parameter is not identifiable, the method
will not claim significance. Hence, some robustness against
nonidentifiability is offered with such a method.

%pa2.4.subsubsection.3 #&#
\textit{Sparsity.}
All the described methods require some sparsity assumption of the parameter
vector $\beta^0$ [if the model is misspecified, this concerns the parameter
$\beta^0$ as in (\ref{betaproj}) or the basis pursuit solution]; see the
discussion of (A1) after Fact~\ref{th1} or assumption (B1). Such sparsity
assumptions can be somewhat relaxed to require weak sparsity in terms of
$\|\beta^0\|_r$ for some $0 < r < 1$, allowing that many or all regression
parameters are nonzero but sufficiently small (cf. \cite{vdg15}; \citep{pbvdg15}).

When the truth (or the linear approximation of the true model) is
nonsparse, the methods are expected to break down. With the Multi
sample-splitting procedure, however, a violation of sparsity might be
detected,
since for nonsparse problems, a sparse variable screening method will be
typically unstable with the consequence that the resulting aggregated
$p$-values are typically not small; see also Section~\ref{subsec.othersparsemeth}.

Finally, we note that for the desparsified Lasso, the sparsity assumption
(B3) or its weaker version can be dropped when using the square root Lasso;
see the discussion after Fact~\ref{th2}.

%pa2.4.subsubsection.4 #&#
\textit{Hidden variables}.
The problem of hidden variables is most prominent in the area of causal
inference (cf. \cite{pearl00}). In the presence of hidden variables, the
presented techniques need to be adapted, adopting ideas from, for
example, the
framework of EM-type estimation (cf. \cite{dempster1977maximum}), low-rank
methods (cf. \cite{chandrasekaran2012}) or the FCI technique from causal
inference (cf. \cite{sgs00}).

%s2.5 #&#
\subsection{A Broad Comparison}\label{subsec.comparlm}
We compare a variety of methods on the basis of multiple testing corrected
$p$-values and single testing confidence intervals. The methods we look at
are the multiple sample-splitting method \emph{MS-Split} (Section~\ref{subsec.multisample-split}), the desparsified Lasso method
\emph{Lasso-Pro} (Section~\ref{subsec.desparslasso}), the Ridge
projection method \emph{Ridge} (Section~\ref{subsec.ridge-proj}), the covariance test \emph{Covtest} (Section~\ref{subsec.othermeth}), the method by Javanmard and Montanari
\emph{Jm2013} (Section~\ref{subsec.othermeth}) and the two bootstrap procedures mentioned in Section~\ref{subsec.othermeth} [\emph{Res-Boot} corresponds to
\citet{chatter13} and \emph{liuyu} to \citet{liuyu13}].

%s2.5.1 #&#
\subsubsection{Specific details for the methods}

For the estimation of the error variance, for the Ridge projection or the
desparsified Lasso method, the scaled Lasso is used as mentioned in
Section~\ref{subsec.addissues}.

For the choice of tuning parameters for the nodewise Lasso regressions
(discussed in Section~\ref{subsec.desparslasso}), we look at the two
alternatives of
using either cross-validation or our more favored alternative procedure (denoted
by Z\&Z) discussed in Appendix~\ref{subsec.appadd}.

We do not look at the bootstrap procedures in connection with multiple
testing adjustment due to the fact that the required
number of bootstrap samples grows out of proportion to go far enough in the
tails of the distribution; some additional importance sampling might help
to address such issues.

Regarding the covariance test, the procedure does not directly provide
$p$-values for the
hypotheses we are interested in. For the sake of comparison though, we use
the interpretation as in \citet{covtestpblmvdg14}.

This interpretation does not have a theoretical reasoning behind it and
functions more as a heuristic.

Thus, the results of the covariance test
procedure should be interpreted with caution.

For the method \emph{Jm2013}, we used our own implementation instead of the
code provided by the authors. The reason for this is that we had already
implemented our own version when we discovered that code was available and
our own version was (orders of magnitude) better in terms of error
control. Posed with the dilemma of fair comparison, we stuck to the best
performing alternative.

%s2.5.2 #&#
\subsubsection{Data used}\label{subsubsec.data}
For the empirical results, simulated design matrices as well as design
matrices from real data are used. The simulated design matrices are
generated $\sim\mathcal{N}_p(0,\Sigma)$ with covariance matrix $\Sigma
$ of
the following three types:
\begin{eqnarray*}
&&\mbox{Toeplitz:}\quad \Sigma_{j,k} = 0.9^{|j-k|},
\\
&&\mbox{Exp.decay:}\quad \bigl(\Sigma^{-1}\bigr)_{j,k} =
0.4^{|j-k|/5},
\\
&&\mbox{Equi.corr:}\quad \Sigma_{j,k} \equiv0.8 \quad\mbox{for all } j \neq k,
\\
&&\hspace*{56pt}\Sigma_{j,j} \equiv1\quad \mbox{ for all } j.
\end{eqnarray*}
The sample size and dimension are fixed at $n=100$ and $p=500$,
respectively. We note that the Toeplitz type has a banded inverse
$\Sigma^{-1}$,
and, vice-versa, the Exp.decay type exhibits a banded $\Sigma$.
The design matrix RealX from real gene expression data of Bacillus Subtilis
($n=71,p=4088$) was
kindly provided by DSM (Switzerland) and is publicly available
(\cite{bumeka13}). To make the
problem somewhat comparable in difficulty to the simulated designs, the
number of variables is reduced to $p=500$ by taking the variables with
highest empirical variance.

The cardinality of the active set is picked to be one of two levels $s_0
\in\{3,15\}$.

For each of the active set sizes, we look at 6 different ways of picking
the sizes of the nonzero coefficients:
\begin{eqnarray*}
&&\mbox{Randomly generated}:\quad U(0,2), U(0,4), U(-2,2),
\\
&&\mbox{A fixed value}:\quad 1, 2 \mbox{ or } 10.
\end{eqnarray*}

The positions of the nonzero coefficients as columns of the design
$\mathbf X$ are picked at random. Results where the nonzero
coefficients were positioned to be the first $s_0$ columns of
$\mathbf X$ can be found in the supplemental article
(\cite{supplement}).

Once we have the design matrix $\mathbf X$ and coefficient vector
$\beta^0$,
the responses $Y$ are generated according to the linear model equation with
$\eps\sim\mathcal{N}(0,1)$.

%f2 #&#
\begin{figure*}

\includegraphics{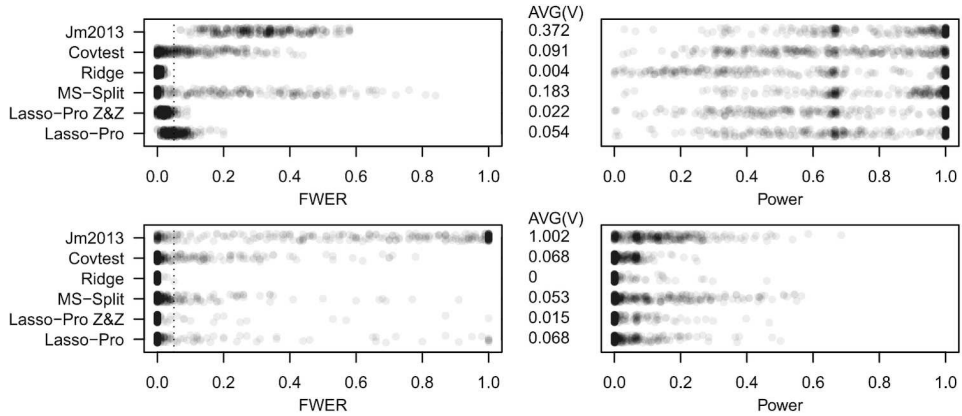}

\caption{Familywise error rate (FWER), average number of false
positive [AVG(V)] and power
for multiple testing based on various methods for a linear model. The
desired control
level for the FWER is
$\alpha=0.05$. The average number of false positives AVG(V) for each
method is shown in the middle. The design matrix is of type
\emph{Toeplitz}, and the active set size being
$s_0=3$ (top) and $s_0=15$ (bottom).}
\label{fig:lintoeplitz}
\end{figure*}

%f3 #&#
\begin{figure*}[b]

\includegraphics{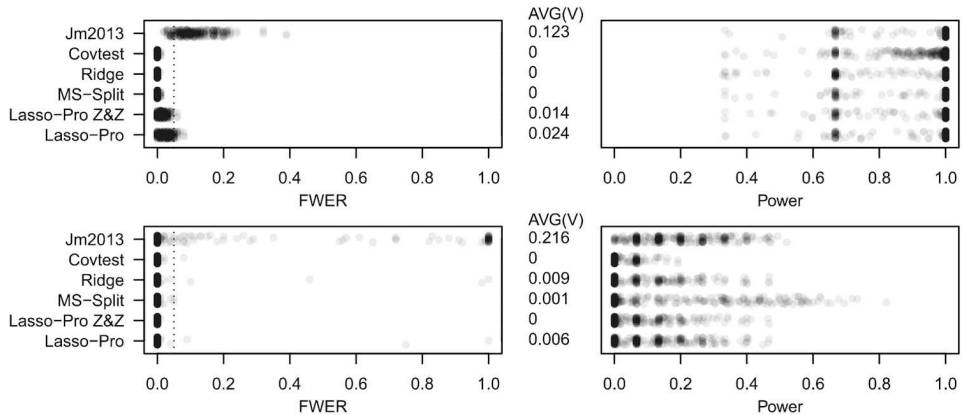}

\caption{See caption of Figure \protect\ref{fig:lintoeplitz} with the only
difference being the type of design matrix. In this plot, the design
matrix type is \emph{Exp.decay}.}
\label{fig:linexpdecay}
\end{figure*}

%s2.5.3 #&#
\subsubsection{$p$-values}\label{subsubsec.pvals}
We investigate multiple testing corrected $p$-values for two-sided
testing of the null hypotheses $H_{0,j}: \beta^0_j = 0$ for $j=1,\ldots
,p$.
We report the power and the familywise error rate (FWER) for each method:
\begin{eqnarray*}
\mbox{Power}& =& \sum_{j \in S_0} \PP[H_{0,j}\mbox{
is rejected}]/s_0,
\\
\mbox{FWER} &=& \PP\bigl[\exists j \in S_0^c :
H_{0,j}\mbox{ is rejected}\bigr].
\end{eqnarray*}
We calculate
%Each data point will represent the result of the calculation of the
empirical versions of these quantities based on fitting 100 simulated
responses $Y$ coming from newly generated $\eps$.

For every design type, active set size and coefficient type combination we
obtain 50 data points of the empirical versions of ``Power'' and ``FWER,''
from 50 independent simulations. Thereby, each data point has a newly generated
$X$, $\beta^0$ (if not fixed) and active set positions $S_0 \in\{1,\ldots,
p\}$; thus, the 50 data points indicate the variability with respect to the
three quantities in the data generation (for the same covariance model of
the design, the same model for the regression parameter and its active set
positions). The data points are grouped in plots by design type and active
set size.

We also report the average number of false positives \texttt{AVG(V)} over
all data points per method next to the FWER plot.

\begin{figure*}

\includegraphics{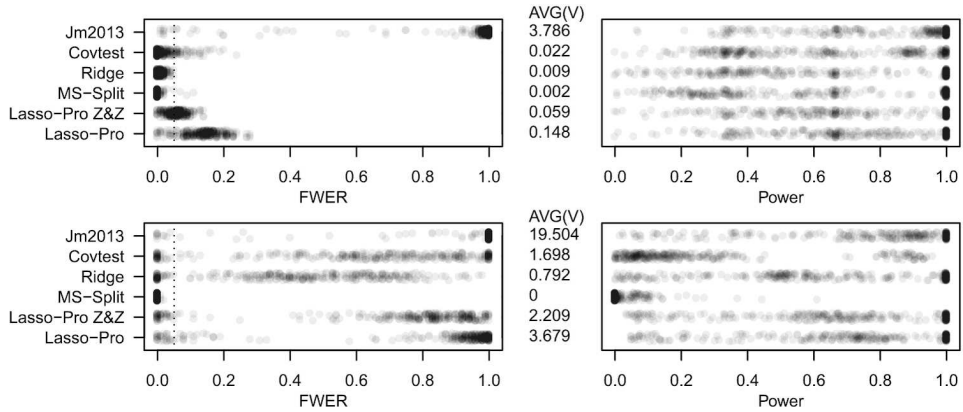}

\caption{See caption of Figure \protect\ref{fig:lintoeplitz} with the only
difference being the type of design matrix. In this plot, the design
matrix type is \emph{Equi.corr}.}
\label{fig:linequi}
\end{figure*}

\begin{figure*}[b]

\includegraphics{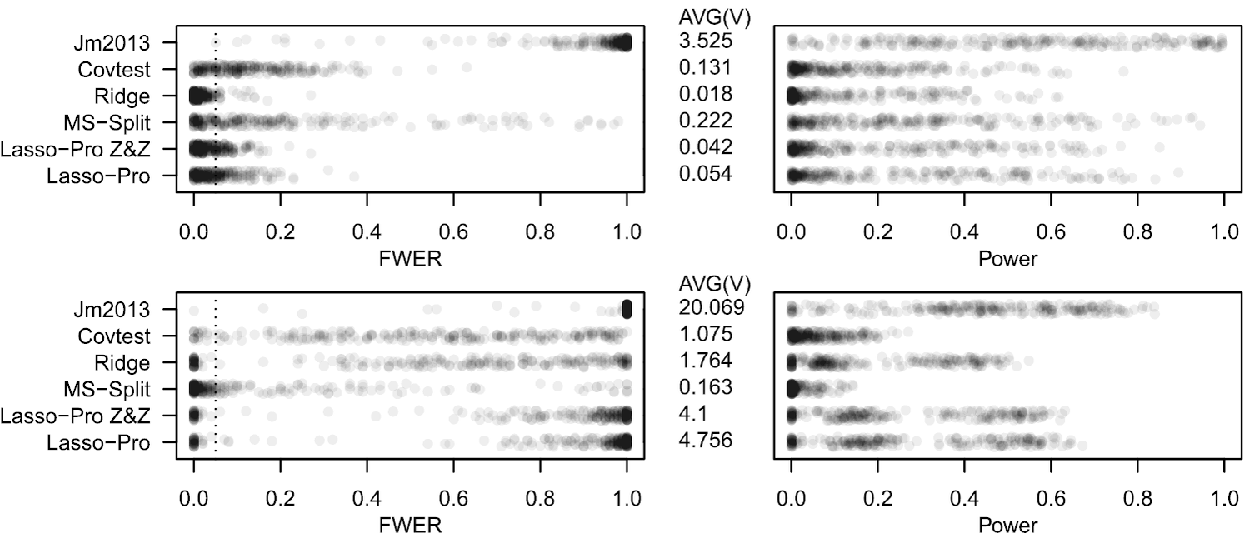}

\caption{See caption of Figure \protect\ref{fig:lintoeplitz} with the only
difference being the type of design matrix. In this plot, the design
matrix type is \emph{RealX}.}
\label{fig:linrealx}
\end{figure*}

The results, illustrating the performance for various methods,
can be found in Figures~\ref{fig:lintoeplitz},
\ref{fig:linexpdecay}, \ref{fig:linequi} and \ref{fig:linrealx}.

%explain multiple pulled setups

%f4 #&#

%f5 #&#

%f6 #&#
\begin{figure*}

\includegraphics{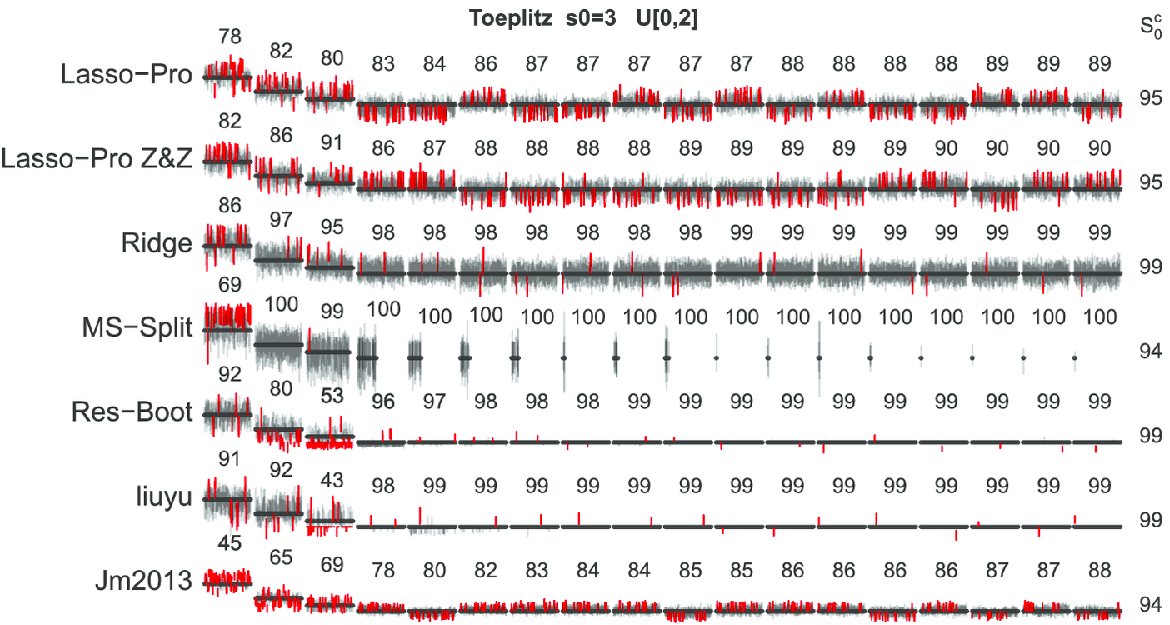}

\caption{Confidence intervals and their coverage rates for 100 realizations of
a linear model with fixed design of dimensions $n=100$, $p=500$. The
design matrix was of type Toeplitz and the active set was of size
$s_0=3$. The nonzero coefficients were chosen by sampling once from
the uniform distribution $U[0,2]$. For each method, 18 coefficients are
shown from left to right with the 100 estimated 95\%-confidence
intervals drawn for each coefficient.The first 3 coefficients are the
non-zero coefficients in descending order of value. The other 15
coefficients, to the right of the first 3, were chosen to be those
coefficients with the worst coverage. The size of each coefficient is
illustrated by the height of a black horizontal bar. To illustrate the
coverage of the confidence intervals, each confidence interval is
either colored red or black depending on the inclusion of the true
coefficient in the interval. Black means the true coefficient was
covered by the interval. The numbers written above the coefficients
are the number of confidence intervals, out of 100, that covered the
truth. All confidence intervals are on the same scale such that one
can easily see which methods have wider confidence intervals. To
summarize the coverage for all zero coefficients $S_0^c$ (including
those not shown on the plot), the rounded average coverage of those
coefficients is given to the right of all coefficients.}
\label{fig:lincitoeplitz}
\end{figure*}

%s2.5.4 #&#
\subsubsection{Confidence intervals}
We investigate confidence intervals for the one particular setup of the
Toeplitz design, active set size $s_0=3$ and coefficients $\beta^0_j
\sim
U[0,2]\ (j \in S_0)$. The active set positions are chosen to be the first
$s_0$ columns of $\mathbf X$. The results we show will correspond
to a
single data point in the $p$-value results.

In Figure~\ref{fig:lincitoeplitz}, 100 confidence intervals are plotted for
each coefficient for each method. These confidence intervals are the
results of fitting 100 different responses Y resulting from newly generated
$\eps$ error terms.

For the Multi sample-splitting method from Section~\ref{subsec.multisample-split}, if a variable did not get selected often
enough in the sample splits, there is not enough information to draw a
confidence interval for it. This is represented in the plot by only drawing
confidence intervals when this was not the case. If the (uncheckable) beta-min
condition (\ref{beta-min}) would be fulfilled, we would know that those
confidence intervals cover zero.

For the bootstrapping methods, an invisible confidence
interval is the result of the coefficient being set to zero in all
bootstrap iterations.

%s2.5.5 #&#
\subsubsection{Summarizing the empirical results}
As a first observation, the impact of the sparsity of the problem on
performance cannot be denied. The power clearly gets worse for $s_0=15$ for
the Toeplitz and Exp.decay setups. The FWER becomes too high for quite a
few methods for $s_0=15$ in the cases of Equi.corr and RealX.

For the sparsity $s_0=3$, the Ridge projection method manages to control
the FWER as desired for all setups. In the case of $s_0=15$, it is the
Multi sample-splitting method that comes out best in comparison to the
other methods. Generally speaking, good error control tends to be
associated with a
lower power, which is not too surprising since we are dealing with the
trade-off between type I and type II errors. The desparsified Lasso method
turns out to be a less conservative alternative with not perfect but
reasonable FWER control as long as the problem is sparse
enough ($s_0=3$). The method has a slightly too high
FWER for the Equi.corr and RealX setups, but FWER around 0.05
for Toeplitz and Exp.decay designs. Doing the Z\&Z tuning procedure
helps the error control, as can be seen most clearly in the Equi.corr
setup. %generally: lasso pro vs lasso pro Z&Z

The results for the simulations where the positions for the nonzero
coefficients were not randomly chosen, presented in the supplemental
article (\cite{supplement}), largely give the same
picture. In comparison to the results presented before,
the Toeplitz setup is easier while the Exp.decay setup is
more challenging. The Equi.corr results are very similar to the ones
from before, which is to be expected from the covariance structure.

Looking into the confidence interval results, it
is clear that the confidence intervals of the Multi sample-splitting
method and the Ridge projection method are wider than the rest.
For the bootstrapping methods, the super-efficiency phenomenon
mentioned in Section~\ref{subsec.othermeth} is visible. Important to
note here is that the smallest nonzero coefficient, the third
column, has very poor coverage from these methods.

We can conclude that the coverage of the zero coefficients is decent
for all methods and that the coverage of the nonzero coefficients is
in line with the error rates for the $p$-values.

Confidence interval results for many other setup combinations are provided
in the supplemental article (\cite{supplement}). The observations are
to a large extent the same.

%s3 #&#
\section{Generalized Linear Models}\label{sec.GLM}

Consider a generalized linear model
\begin{eqnarray*}
& &Y_1,\ldots,Y_n\quad \mbox{independent},
\\
& &g\bigl(\EE[Y_i|X_i = x]\bigr) = \mu^0 +
\sum_{j=1}^p \beta^0_j
x^{(j)},
\end{eqnarray*}
where $g(\cdot)$ is a real-valued, known link function. As before, the goal
is to construct confidence intervals and statistical tests for the unknown
parameters $\beta^0_1,\ldots,\beta^0_p$, and maybe $\mu^0$ as well.

%s3.1 #&#
\subsection{Methods}\label{subsec.GLMmethods}

The Multi sample-splitting method can be modified for GLMs in an obvious
way: the variable screening step using the first half of the data can be
based on the $\ell_1$-norm regularized MLE, and $p$-values and confidence
intervals using the second half of the sample are constructed from the
asymptotic distribution of the (low-dimensional) MLE. Multiple testing
correction and aggregation of the $p$-values from multiple sample splits are
done exactly as for linear models in Section~\ref{subsec.multisample-split}.

A desparsified Lasso estimator for GLMs can be constructed as follows
(\cite{vdgetal13}): The
$\ell_1$-norm regularized MLE $\hat{\theta}$ for the parameters $\theta
^0 =
(\mu^0,\beta^0)$ is desparsified with a method based on the
Karush--Kuhn--Tucker (KKT) conditions for $\hat{\theta}$, leading to an
estimator with an asymptotic Gaussian distribution. The Gaussian
distribution can then be used to construct confidence intervals and
hypothesis tests.
%s3.2 #&#
\subsection{Weighted Squared Error Approach}\label{subsec.GLMweighted}

The problem can be simplified in such a way that we can apply the
approaches for the linear model from Section~\ref{sec.LM}. This can be done
for all types of generalized linear models (as shown in Appendix~\ref{subsec.app.general.wsqerr}), but we restrict ourselves in this section
to the specific case of logistic regression. Logistic regression is
usually fitted
by applying the iteratively reweighted least squares (IRLS)
algorithm where at every iteration one solves a weighted least squares
problem (\cite{hastetal09}).

The idea is now to apply a standard l1-penalized fitting of the model, build
up the weighted least squares problem at the l1-solution and then apply
our linear model methods on this problem.

We use the notation $\hat{\pi}_i, i = 1, \ldots,n$ for the
estimated probability of the binary outcome. $\hat{\pi}$ is the vector of
these probabilities.

%f7 #&#
\begin{figure*}

\includegraphics{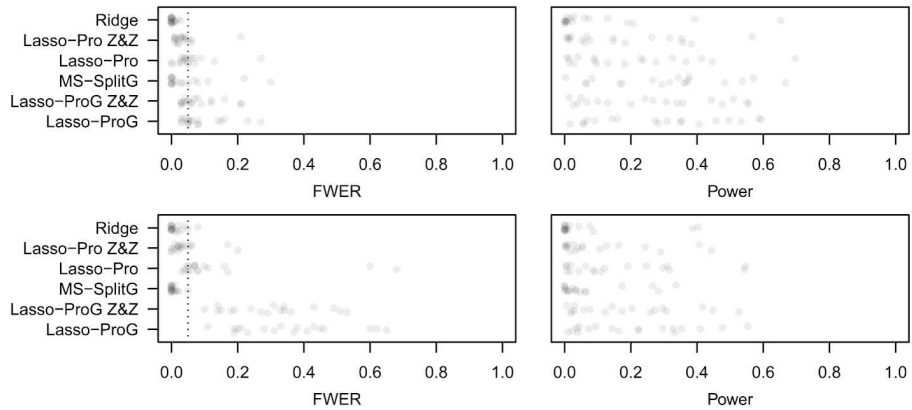}

\caption{Familywise error rate (FWER) and power
for multiple testing based on various methods for logistic regression.
The desired control
level for the FWER is $\alpha=0.05$. The design matrix is of type
\emph{Toeplitz} in the top plot and \emph{Equi.corr} in the bottom
plot. If the method name contains a capital \texttt{G}, it is the
modified glm version, otherwise the linear model methods are using
the weighted squared error approach.}
\label{fig:glmsimul}
\end{figure*}

From \citet{hastetal09}, the adjusted response variable becomes
\[
Y_{\mathrm{adj}} = \mathbf X \hat{\beta} + \mathbf W^{-1}(Y-\hat{
\pi}),
\]
and the weighted least squares problem is
\[
\hat{\beta}_{\mathrm{new}} = \argmin_{\beta} (Y_{\mathrm{adj}} - \mathbf
X \beta)^T \mathbf W (Y_{\mathrm{adj}} - \mathbf X \beta),
\]
with weights
\[
\mathbf W = %
\pmatrix{ \hat{\pi}_1(1-\hat{
\pi}_1) & 0 & \ldots& 0\vspace*{2pt}
\cr
0 & \hat{
\pi}_2(1-\hat{\pi}_2) & \ddots& \vdots\vspace*{2pt}
\cr
\vdots& \ddots& \ddots& 0\vspace*{2pt}
\cr
0 & \ldots& 0 & \hat{
\pi}_n(1-\hat{\pi}_n) } %
\hspace*{-0.5pt}.
\]

We rewrite $Y_{w} = \sqrt{\mathbf W} Y_{\mathrm{adj}}$ and $X_w =
\sqrt{\mathbf W} \mathbf X$ to get
\[
\hat{\beta}_{\mathrm{new}} = \argmin_{\beta} (Y_w - \mathbf
X_w \beta)^T(Y_w - \mathbf X_w
\beta).
\]

The linear model methods can now be applied to $Y_{w}$ and
$\mathbf X_{w}$, thereby the estimate $\hat{\sigma}_{\eps}$ has to
be set to the value
1. We note that in the low-dimensional case, the resulting $p$-values (with
unregularized residuals $Z_j$) are very similar to the $p$-values
provided by
the standard \texttt{R}-function \texttt{glm}.

%s3.3 #&#
\subsection{Small Empirical Comparison}

We provide a small empirical comparison of the methods mentioned in
Sections~\ref{subsec.GLMmethods} and \ref{subsec.GLMweighted}. When applying the linear
model procedures, we use the naming from Section~\ref{subsec.comparlm}. The new
GLM-specific methods from Section~\ref{subsec.GLMmethods} are referred to by their
linear model names with a capital G added to them.

For simulating the data, we use a subset of the variations presented in Section~\ref{subsubsec.data}.
We only look at Toeplitz and Equi.corr and an active set size of
$s_0=3$. The number of variables is fixed at $p=500$, but the sample
size is
varied $n\in\{100,200,400\}$.
The coefficients were randomly generated:
\begin{eqnarray*}
\mbox{Randomly generated}:\quad U(0,1), U(0,2), U(0,4).
\end{eqnarray*}
The nonzero coefficient positions are chosen randomly in one case and
fixed as the first $s_0$ columns of $\mathbf X$ in the other.

For every combination (of type of design, type of coefficients,
sample size and coefficient positions), 100 responses $Y$ are simulated to
calculate empirical versions of the ``Power'' and ``FWER'' described in
Section~\ref{subsubsec.pvals}.
In contrast to the $p$-value results from Section~\ref{subsubsec.pvals},
there is only one resulting data point per setup combination (i.e., no
additional replication with new random covariates, random coefficients and
random active set). For each
method, there are 18 data points, corresponding to 18 settings, in each plot.
The results can be found in Figure~\ref{fig:glmsimul}.

Both the modified GLM methods as well as the weighted squared error
approach work adequately. The Equi.corr setup does prove to be
challenging for \emph{Lasso-ProG}.

%s3.4 #&#
\subsection{\texttt{hdi} for Generalized Linear Models}
In the \texttt{hdi} \textsf{R}-package (\cite{hdipackage}) we also provide
the option to use the Ridge projection method and the desparsified Lasso
method with the weighted squared error approach.

We provide the option to specify the \texttt{family} of the response
$Y$ as
done in the \textsf{R}-package \texttt{glmnet}:

\begin{verbatim}
> outRidge
  <- ridge.proj(x = x, y = y,
     family = ''binomial'')
> outLasso
  <- lasso.proj(x = x, y = y,
     family = ''binomial'')
\end{verbatim}

$p$-values and confidence intervals are extracted in the exact same way
as for the linear model case; see Section~\ref{subsec.hdilin}.

%s4 #&#
\section{Hierarchical Inference in the Presence of Highly Correlated
Variables}\label{sect.hierinf}

The previous sections and methods assume in some form or another that
the effects are strong enough to enable accurate estimation of the
contribution of \emph{individual variables}.

Variables are often highly correlated for high-dimensional
data. Working with a small sample size, it is impossible to attribute
any effect to
an individual variable if the correlation between a block of variables
is too high. Confidence intervals for individual
variables are then very wide and uninformative. Asking for confidence
intervals for individual variables thus leads to poor power of all
procedures considered so far. Perhaps even worse, under high correlation
between variables the coverage of some procedures will also be
unreliable as the necessary conditions for correct coverage (such as
the compatibility assumption) are violated.

In such a scenario, the individual effects are not granular enough
to be resolved. However, it might yet still be possible to attribute an
effect to a group
of variables. The groups can arise naturally due to a specific
structure of
the problem, such as in applications of the \emph{group
Lasso} (\cite{yuan06}).

Perhaps more often, the groups are derived
via hierarchical clustering (\cite{hartigan1975clustering}), using the
correlation structure or some
other distance between the variables.
The main idea is as
follows. A hierarchy ${\cal T}$ is a set
of clusters or groups $\{{\cal C}_k; k\}$ with ${\cal C}_k \subseteq
\{1,\ldots,p\}$. The root node (cluster) contains all variables
$\{1,\ldots,p\}$. For any two clusters ${\cal C}_k, {\cal C}_{\ell}$,
either one cluster is a subset of the other or they have an empty
intersection. Usually, a hierarchical clustering has an additional notion
of a level such that, on each level, the corresponding clusters build a
partition of $\{1,\ldots,p\}$. We consider a hierarchy ${\cal T}$ and
first test the root node cluster
${\cal C}_0
= \{1,\ldots,p\}$ with
hypothesis $H_{0,{\cal C}_0}: \beta_1 = \beta_2 = \cdots= \beta_p =
0$. If this hypothesis is rejected, we test the next clusters ${\cal C}_k$
in the hierarchy (all clusters whose supersets are the root node
cluster ${\cal
C}_0$ only): the corresponding cluster hypotheses are $H_{0,{\cal C}_k}:
\beta_j = 0$ for all $j \in{\cal C}_k$. For the hypotheses which can be
rejected, we consider all smaller clusters whose only supersets are
clusters which have been rejected by the method before, and we continue to
go down the tree hierarchy until no more cluster hypotheses can be
rejected.

With the hierarchical scheme in place, we still need a test for the
null hypothesis $H_{0,{\cal C}}$ of a cluster of variables. The tests
have different properties. For example, whether a multiplicity
adjustment is necessary will depend on the chosen test.
We will describe below some methods that are useful for testing
the effect of a group of variables and which can be used in such a
hierarchical approach. The
nice and interesting feature of the procedures is that they adapt
automatically to the level of the hierarchical tree: if a signal of a small
cluster of variables is strong, and if that cluster is sufficiently uncorrelated
from all other variables or clusters, the cluster will be detected as
significant.
Vice-versa, if the signal is weak or if the cluster has too high a
correlation with other variables or clusters, the cluster will not
become significant. For example, a single variable
cannot be detected as significant if it has too much correlation to
other variables or clusters.

%s4.1 #&#
\subsection{Group-Bound Confidence Intervals Without Design
Assumptions}\label{subsec.assfree}

The \emph{Group-bound} proposed in \citet{meins13} gives confidence
intervals for the $\ell_1$-norm $\|\beta^0_{{\cal C}_k}\|_1$ of a
group ${{\cal C}_k}\subseteq\{1,\ldots,p\}$ of variables. If the
lower-bound of the $1-\alpha$ confidence interval is larger than 0,
then the null hypothesis $\beta^0_{{\cal C}_k}\equiv0$ can be rejected
for this group. The method combines a few properties:
\begin{longlist}[(iii)]
\item[(i)] The confidence intervals are valid without an assumption
like the compatibility condition (\ref{compat}). In general, they
are conservative, but if the compatibility condition holds, they have
good ``power'' properties (in terms of length) as well.
\item[(ii)] The test is hierarchical. If a set of variables can be
rejected, all
supersets will also be rejected. And vice-versa, if a group of
variables cannot be rejected, none of its subsets can be rejected.
\item[(iii)] The estimation accuracy has an optimal detection rate
under the so-called group effect compatibility condition,
which is weaker than the compatibility condition necessary to
detect the effect of individual variables.
\item[(iv)] The power of the test is unaffected by adding highly or
even perfectly correlated variables in ${\cal C}_k $ to the
group. The compatibility condition would fail to yield a
nontrivial bound, but the group effect compatibility
condition is unaffected by the addition of perfectly correlated
variables to a group.
\end{longlist}
The price to pay for the assumption-free nature of the bound is a weaker
power than with previously discussed approaches when the goal is to detect
the effect of individual variables. However, for groups of highly
correlated variables, the approach can be much more powerful than simply
testing all variables in the group.

%
%f8 #&#
\begin{figure*}

\includegraphics{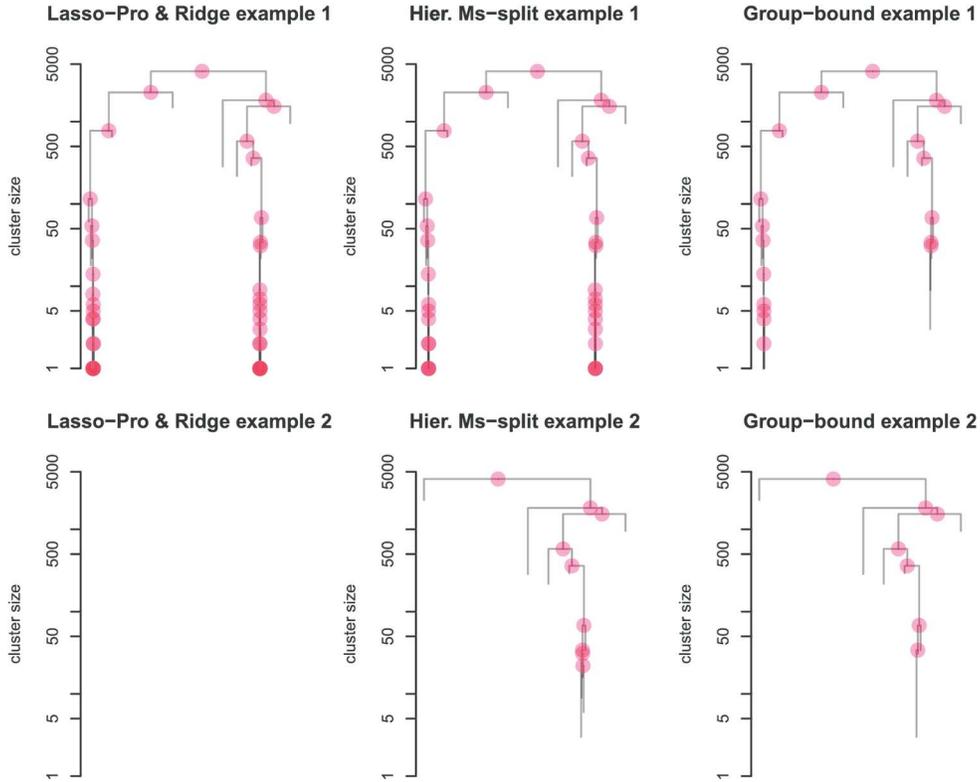}

\caption{A visualization of the hierarchical testing
scheme as described in the beginning of Section~\protect\ref{sect.hierinf}, for the examples described in
Section~\protect\ref{subsec.illustrations}. One moves top-down through
the output of a hierarchical clustering scheme, starting at the root
node. For each cluster encountered, the null hypothesis that all the
coefficients
of that particular cluster are 0 is tested. A rejection is visualized
by a red
semi-transparent circle at a vertical position that corresponds to
the size of the cluster. The chosen significance level was $\alpha=0.05$.
The children of significant clusters in the
hierarchy are connected by a black line. The process is repeated by
testing the null hypotheses for all
those children clusters until no more hypotheses could
be rejected.
The ordering of the hierarchy in the horizontal direction has
no meaning and was chosen for a clean separation of children hierarchies.
The hierarchical clustering and orderings are the same for all 6
plots since the design matrix was the same. Two different examples
were looked at (corresponding to top and bottom row, resp.) and
four different methods were applied to these
examples. The desparsified Lasso and the Ridge method gave identical
results and were grouped in the two plots on the left, while
results from the hierarchical Multi sample-splitting method are
presented in the middle column and the results for the Group-bound
method are
shown in the right column. In example 1,
the responses were simulated with 2 clusters of
highly correlated variables of size 3 having coefficients different
from zero. In example 2, the responses were
simulated with 2 clusters of highly correlated variables of sizes 11
and 21 having coefficients different from zero. More details about
the examples can be found in Section \protect\ref{subsec.illustrations}.}
\label{fig:treeridge}
\end{figure*}

We remark that previously developed tests can be adapted to the context of
hierarchical testing of groups with hierarchical adjustment for
familywise error control
(\cite{Meins08}); for the Multi sample-splitting
method, this is described next.

%s4.2 #&#
\subsection{Hierarchical Multi Sample-Splitting}\label{subsec.mssplitgroup}

The Multi sample-splitting method (Section~\ref{subsec.multisample-split}) can
be adapted to the context of
hierarchical testing of groups by using hierarchical adjustment of\vadjust{\goodbreak}
familywise error control (\cite{Meins08}).
When testing a cluster hypotheses $H_{0,{\cal C}}$, one can use a modified
form of the
partial $F$-test for high-dimensional settings; and the multiple testing
adjustment due to the multiple cluster hypotheses considered can be taken
care of by a hierarchical adjustment scheme proposed in \citet
{Meins08}. A
detailed description of the method, denoted here by \emph{Hier. MS-Split},
together with theoretical guarantees is given in \citet{manbu13}.

%s4.3 #&#
\subsection{Simultaneous Inference with the Ridge or Desparsified Lasso
Method}\label{subsec.simulcovridgelasso}

Simultaneous inference for all possible groups can be achieved by considering
$p$-values $P_j$ of individual hypotheses $H_{0,j}: \beta^0_j = 0$
($j=1,\ldots,p$) and adjusting them for simultaneous coverage, namely,
$P_{\mathrm{adjusted},j} = P_j \cdot p$. The individual $p$-values $P_j$ can
be obtained by the Ridge or desparsified Lasso method in
Section~\ref{sec.LM}.

We can then test any group hypothesis $H_{0,G}: \beta_j^0 = 0$ for all $j
\in G$ by simply looking whether $\min_{j \in G} P_{\mathrm{adjust},j}
\le
\alpha$, and we can consider as many group hypotheses as we want without
any further multiple testing adjustment.

%s4.4 #&#
\subsection{Illustrations}\label{subsec.illustrations}

A semi-real data example is shown in Figure~\ref{fig:treeridge}, where the
predictor variables are taken from the Riboflavin data set (\cite{bumeka13})\vadjust{\goodbreak}
($n=71, p=4088$) and the
coefficient vector is taken to have entries 0,
except for 2 clusters of highly correlated variables. In example 1, the
clusters both have size 3 with nonzero coefficient sizes equal to 1 for all
the variables in the clusters and Gaussian noise level
$\sigma=0.1$. In example 2, the clusters are bigger and have different sizes
11 and 21; the coefficient sizes for all the variables in the clusters is
again 1, but the Gaussian noise level here is
chosen to be $\sigma=0.5$.

In the first example, 6 out of the 6 relevant
variables are discovered as individually significant by the
\emph{Lasso-Pro}, \emph{Ridge} and \emph{MS-Split} methods (as outlined in
Sections~\ref{subsec.multisample-split}--\ref{subsec.desparslasso}),
after adjusting for
multiplicity.

In the second example, the methods cannot reject the single variables
individually any longer. The results for the \emph{Group-bound} estimator
are shown in the right column. The \emph{Group-bound} can reject a group
of 4 and 31 variables in the first example, each containing a true
cluster of
3 variables. The method can also detect a group of 2 variables (a
subset of
the cluster of~4) which contains 2 out of the 3 highly correlated
variables. In the second example, a group of 34
variables is rejected with the \emph{Group-bound} estimator, containing 16
of the group of 21 important variables. The smallest group of variables
containing the cluster of 21 that the method can detect is of size 360. It
can thus be detected that the variables jointly have a substantial
effect even
though the null hypothesis cannot be rejected for any variable
individually. The hierarchical Multi sample-splitting method (outlined in
Section~\ref{subsec.mssplitgroup}) manages to detect the same clusters as
the \emph{Group-bound} method. It even goes one step further by
detecting a
smaller subcluster.

%f9 #&#
\begin{figure}

\includegraphics{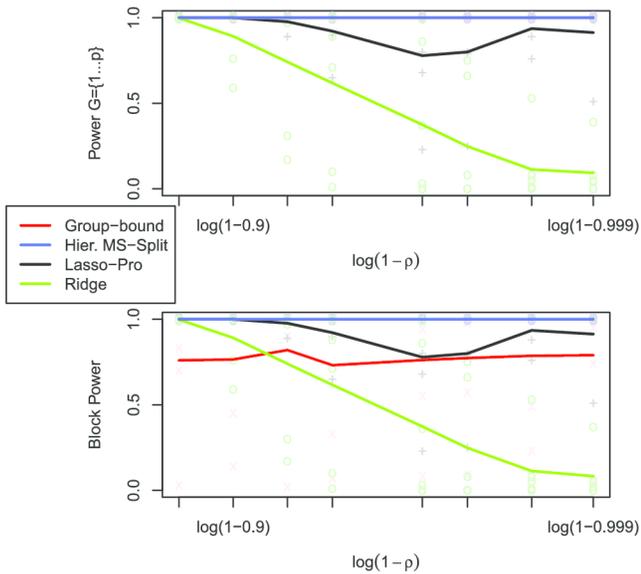}

\caption{The power for the rejection of the group-hypothesis of all
variables (top) and the power for the rejection of the group-hypothesis
of the variables
in blocks highly correlated with $S_0$ variables (bottom). The design
matrix used is of type \emph{Block Equi.corr} which is similar to the
Equi.corr setup in that $\Sigma$ is block diagonal with blocks (of size
$20 \times20$) being the~$\Sigma$ of Equi.corr. The power is plotted
as a function of the correlations in the blocks, quantified by~$\rho$.
The Ridge-based method loses power as the correlation between variables
increases, while the group bound, Hier. MS-Split and Lasso-Pro methods
can maintain
power close to 1 for both measures of power.}\vspace*{6pt}
\label{fig:testgrouppower}
\end{figure}

%f10 #&#
\begin{figure}

\includegraphics{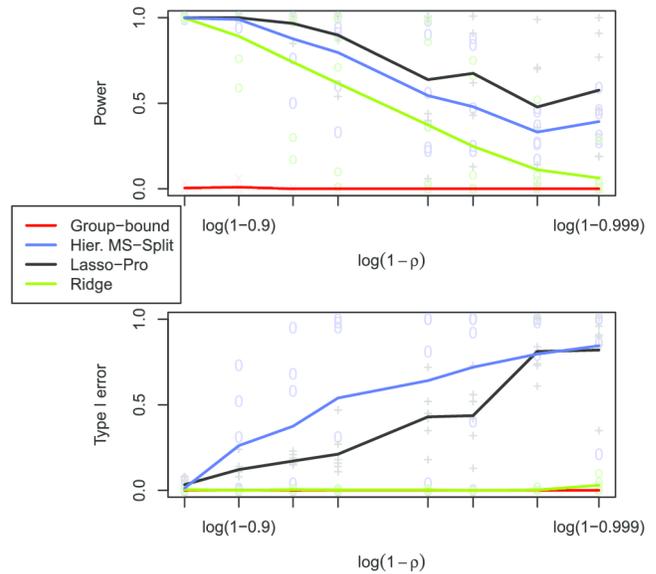}

\caption{The power for the rejection of the group-hypothesis of all $S_0$
variables (top) and type I error rate corresponding to the rejection of
the group-hypothesis of all $S_0^c$ variables (bottom) for the design
matrix of type \emph{Block Equi.corr} when changing the correlation
$\rho$ between variables. The design matrix type is described in detail
in the caption of Figure \protect\ref{fig:testgrouppower} and in the text. The
desparsified Lasso, Hier. MS-Split and the Ridge-based
method lose power as the correlation between variables increases,
while the \emph{Group-bound} cannot reject the small group of
variables $S_0$ (3~in
this case). The desparsified Lasso and MS-Split methods also exceed the
nominal type I error rate for high correlations (as the design
assumptions break down), whereas the Ridge-based method and the
\emph{Group-bound} are both within the nominal 5\% error rate for
every correlation strength.}\vspace*{6pt}
\label{fig:testgroup}
\end{figure}

We also consider the following simulation model. %The design matrix was
%chosen to be Gaussian with mean zero and population block-diagonal
%covariance matrix $\Sigma$ with blocks of dimension $20 \times20$ with
%equi-correlated off-diagonal entries $\rho$. We chose the dimension
%$p=500$, sample size $n=100$ and Gaussian noise with level $\sigma=1$. There
%were only 3 nonzero coefficients chosen with three different signal levels
%(realizations of the $\beta$'s) from Uniform distributions
%$U[0,2]$, $U[0,4]$ and $U[0,8]$. Aside from varying signal
%level, we studied two cases where either all the nonzero
%coefficients were contained in one single highly correlated block or each
%of the active variables were in a different block. Figures~\ref{fig:testgrouppower} and \ref{fig:testgroup} show the coverage
%and power as a function of the correlation $\rho$ between variables in
%another simulation model.
The type
of design matrix was chosen to be such that the
population covariance matrix $\Sigma$ is a block-diagonal matrix with
blocks of dimension $20 \times20$ being of the same type as $\Sigma$
for Equi.corr (see
Section~\ref{subsubsec.data}) with off-diagonal $\rho$ instead of
$0.8$. The dimensions of the problem were chosen to be $p=500$ number of
variables, $n=100$ number of samples and noise level $\sigma=1$. There
were only 3 nonzero coefficients chosen
with three different signal levels $U[0,2]$, $U[0,4]$ and $U[0,8]$ being used
for the
simulations. Aside from varying signal
level, we studied the two cases where in one case all the nonzero
coefficients were contained in one single highly correlated block and in
the other case each of those variables was in a different block.\vadjust{\goodbreak} We
look at
3 different measures of power. One can define the power as the fraction
of the 100
repeated simulations that the method managed to reject the group of all
variables $G =
{1,\ldots,p}$. This is shown at the top in Figure~\ref{fig:testgrouppower}. Alternatively, one can look at the rejection
rate of the hypothesis for the group $G$ that contains all variables in the
highly correlated blocks that contain a variable from $S_0$. This is the
plot at the bottom in Figure~\ref{fig:testgrouppower}.
Finally, one can look at the rejection rate of the hypothesis where the
group $G$ contains only the variables in $S_0$ (of size 3 in this
case). The type I error we define to be the fraction of the
simulations in which the method rejected the group hypothesis $H_{0,S_0^c}$
where all regression coefficients are equal to zero. These last two measures are
presented in Figure~\ref{fig:testgroup}.

The power of the Ridge-based method (\cite{pb13}) drops substantially
for high
correlations. The power of the \emph{Group-bound} stays close to 1
at the level of the highly correlated groups (Block-power) and above (Power
$G={1 ,\ldots, p}$) throughout the entire range of correlation values. The
\emph{Lasso-Pro} and \emph{MS-Split} perform well here as well. The power
of the\vadjust{\goodbreak} \emph{Group-bound} is 0 when attempting to reject the small
groups $H_{0,S_0}$.
The type I error rate is supposedly controlled at level $\alpha=0.05$
with all three
methods. However, the \emph{Lasso-Pro} and the hierarchical \emph
{MS-Split} methods fail
to control the error rates, with the type I error
rate even approaching 1 for large values of the correlation. The
\emph{Group-bound} and Ridge-based estimator have, in contrast, a type I
error rate close to 0 for all values of the correlation.

For highly correlated groups of variables, trying to detect the effect
of individual variables has thus two inherent dangers. The power to
detect interesting groups of variables might be very low. And the
assumptions for the methods might be violated, which invalidates the
type I error control. The assumption-free \emph{Group-bound} method
provides a powerful test for the group effects even if variables are
perfectly correlated, but suffers in power, relatively speaking, when
variables are not highly correlated.

%s4.5 #&#
\subsection{\texttt{hdi} for Hierarchical Inference}
An implementation of the \emph{Group-bound} method is provided in the
\texttt{hdi} \textsf{R}-package (\cite{hdipackage}).

For specific groups, one can provide a vector or a list of vectors where
the elements of the vector specify the desired columns of $\bx$
to be tested for.
The following code tests the group hypothesis if the group contains all
variables:

\begin{verbatim}
> group
  <- 1:ncol(x)
> outGroupBound
  <- groupBound(x = x, y = y,
     group = group, alpha = 0.05)
> rejection
  <- outGroupBound > 0
\end{verbatim}

Note that one needs to specify the significance level~$\alpha$.

One can also let the method itself apply the hierarchical clustering scheme
as described at the beginning of Section~\ref{sect.hierinf}.

This works as follows:

\begin{verbatim}
> outClusterGroupBound
  <- clusterGroupBound(x = x,
     y = y, alpha = 0.05)
\end{verbatim}

The output contains all clusters that were tested for significance in
\texttt{members}. The corresponding lower bounds are contained in
\texttt{lowerBound}.

To extract the significant clusters, one can do

\begin{verbatim}
> significant.cluster.numbers
  <- which
     (outClusterGroupBound
     $lowerBound > 0)
> significant.clusters
  <- outClusterGroupBound$members
     [[significant.cluster.numbers]]
\end{verbatim}

The figures in the style of Figure~\ref{fig:treeridge} can be achieved by
using the function \texttt{plot} on \texttt{outCluster-\break GroupBound}.

Note that one can specify the distance matrix used for the hierarchical
clustering, as done for \texttt{hclust}.

To test group hypotheses $H_{0,G}$ for the Ridge and desparsified Lasso
method as described in Section~\ref{subsec.simulcovridgelasso}, one uses
the output from the original single parameter fit, as illustrated for the
group of all variables:

\begin{verbatim}
> outRidge
  <- ridge.proj(x = x, y = y)
> outLasso
  <- lasso.proj(x = x, y = y)
> group
  <- 1:ncol(x)
> outRidge$groupTest(group)
> outLasso$groupTest(group)
\end{verbatim}

To apply a hierarchical clustering scheme as done in
\texttt{clusterGroupBound}, one calls \texttt{cluster-\break GroupTest}:

\begin{verbatim}
> outRidge$clusterGroupTest
  (alpha = 0.95)
\end{verbatim}

To summarize, the \textsf{R}-package provides functions to test individual
groups as well as to test according to a hierarchical clustering scheme for
the methods \emph{Group-bound}, Ridge and desparsified Lasso.
An implementation of the hierarchical Multi sample-splitting method is not
provided at this point in time.

%s5 #&#
\section{Stability Selection and Illustration with \texttt{hdi}}

Stability selection (\cite{mebu10}) is another methodology to guard against
false positive selections, by controlling the expected number of false
positives $\EE[V]$. The focus is on selection of a single or a group of
variables in a regression model, or on a selection of more general discrete
structures such as graphs or clusters. For example, for a linear model in
(\ref{mod.lin}) and
with a selection of single variables, stability selection provides a subset
of variables $\hat{S}_{\mathrm{stable}}$ such that for $V = |\hat
{S}_{\mathrm{stable}}
\cap S_0^c|$ we have that $\EE[V] \le M$, where $M$ is a prespecified
number.

For selection of single variables in a regression model, the method does
not need a beta-min assumption, but the theoretical analysis of stability
selection for controlling $\EE[V]$ relies on a restrictive exchangeability
condition (which, e.g., is ensured by a restrictive condition on the design
matrix). This exchangeability condition seems far from necessary though
(\cite{mebu10}). A refinement of stability selection is given in
\citet{shah13}.

An implementation of the stability selection procedure is available in the
\texttt{hdi} \textsf{R}-package. It is called in a very similar way as the
other methods. If we want to control, for example, $\EE[V] \le1$, we use

\begin{verbatim}
> outStability
  <- stability
     (x = x, y = y, EV = 1)
\end{verbatim}

The ``stable'' predictors are available in the element \texttt{select}.

The default model selection algorithm is the Lasso (the first $q$
variables entering the Lasso paths).\vadjust{\goodbreak} The option \texttt{model.selector}
allows to apply a user defined model selection function.

%s6 #&#
\section{R Workflow Example}
We go through a possible \texttt{R} workflow based on the Riboflavin
data set
(\cite{bumeka13}) and methods provided in the \texttt{hdi}
\texttt{R}-package:

\begin{verbatim}
> library(hdi)
> data(riboflavin)
\end{verbatim}

We assume a linear model and we would like to investigate which effects are
statistically significant on a significance level of
$\alpha=0.05$. Moreover, we want to construct the corresponding confidence
intervals.

We start by looking at the individual variables.
We want a conservative approach and, based on the
results from Section~\ref{subsec.comparlm}, we choose the Ridge projection
method for its good error control:

\begin{verbatim}
> outRidge
  <- ridge.proj
     (x = riboflavin$x,
      y = riboflavin$y)
\end{verbatim}

We investigate if any of the multiple testing corrected $p$-values are smaller
than our chosen significance level:

\begin{verbatim}
> any(outRidge$pval.corr <= 0.05)
[1] FALSE
\end{verbatim}

We calculate the 95\% confidence intervals for the first 3 predictors:

\begin{verbatim}
> confint(outRidge,parm=1:3,
  level=0.95)
lower upper
AADK_at -0.8848403 1.541988
AAPA_at -1.4107374 1.228205
ABFA_at -1.3942909 1.408472
\end{verbatim}

Disappointed with the lack of significance for testing individual
variables, we want to investigate if we can find a significant group
instead. From the procedure proposed for the Ridge method in Section~\ref{sect.hierinf}, we know that if the Ridge method can not find any
significant individual variables, it would not find a significant group
either.

We apply the Group-bound method with its clustering option to try to find
a significant group:

\begin{verbatim}
> outClusterGroupBound
  <- clusterGroupBound
     (x = riboflavin$x,
     y = riboflavin$y,
     alpha = 0.05)
> significant.cluster.numbers
  <- which(outClusterGroupBound
     $lowerBound
     > 0)
> significant.clusters
  <- outClusterGroupBound
     $members
     [[significant.cluster.numbers]]
> str(significant.clusters)
num [1:4088] 1 2 3 4 5 6 7 8 9 10...
\end{verbatim}

Only a single group, being the root node of the clustering tree, is found
significant.

These results are in line with the results achievable in earlier
studies of
the same data set in \citet{bumeka13} and \citet{vdgetal13}.

%s7 #&#
\section{Concluding Remarks}

We present a (selective) overview of recent developments in frequentist
high-dimensional inference for constructing confidence intervals and
assigning \mbox{$p$-}values for the parameters in linear and generalized linear
models. We include some methods which are able to detect significant groups
of highly correlated variables which cannot
 be individually detected as single
variables. We complement the methodology and theory viewpoints with
a broad empirical study. The latter indicates that more ``stable''
procedures based on Ridge estimation or sample splitting with subsequent
aggregation might be more reliable for type I error control, at the price
of losing power; asymptotically, power-optimal methods perform nicely in
well-posed scenarios but are more exposed to fail for error control in more
difficult settings where
the design or degree of sparsity are more ill-posed. We introduce the
\texttt{R}-package \texttt{hdi} which allows the user to choose from a
collection of frequentist inference methods and eases reproducible
research.

%s7.1 #&#
\subsection{Post-Selection and Sample Splitting
Inference}\label{subsec.postsel}

Since the main assumptions outlined in Section~\ref{subsec.mainass} might
be unrealistic in practice, one can consider a different route.

The
view and ``POSI'' (Post-Selection Inference) method by
\citet{berketal13} makes inferential statements which
are protected against all possible submodels and, therefore, the procedure
is not exposed to
the issue of having selected an ``inappropriate'' submodel. The way in
which \citet{berketal13} deal with misspecification of the (e.g., linear)
model is closely
related to addressing this issue with the Multi sample splitting or
desparsified Lasso method; see Section~\ref{subsec.mainass} and
\citet{pbvdg15}. The method by \citet{berketal13} is conservative, as it
protects against any possible submodel, and it is not feasible yet for
high-dimensional problems. \citet{wass14} briefly describes the ``HARNESS''
(High-dimensional Agnostic Regression Not Employing Structure or Sparsity)
procedure: it is based on single data splitting and making inference for
the selected submodel from the first half of the data. When giving up on
the goal to infer the true or best approximating parameter $\beta^0$ in
(\ref{betaproj}), one can drop many of the main assumptions which are needed
for high-dimensional inference.

The ``HARNESS'' is related to post-selection inference where the
inefficiency of sample splitting is avoided. Some recent work includes
exact post-selection inference, where the full data is used for
selection and inference: it aims to avoid the potential inefficiency of
single sample splitting and to be less conservative than ``POSI'', thereby
restricting the focus to a class of selection procedures which are
determined by
affine inequalities, including the Lasso and least angle regression
(\cite{lee13}; \citep{taylor14}; \citep{fithian14}).

Under some conditions, the issue of selective inference can be
addressed by using an adjustment factor (\cite{beye05}): this could be done
by adjusting the output of our high-dimensional inference procedures,
for example,
from the \texttt{hdi} \texttt{R}-package.

\begin{appendix}\label{app}

%s8 #&#
\section*{Appendix}

%s8.1 #&#
\setcounter{subsection}{0}
\subsection{Additional Definitions and Descriptions}\label{subsec.appadd}

\emph{Compatibility condition} (\cite{pbvdg11}, page106).
Consider a fixed design matrix $\bx$. We define the following:

The compatibility condition holds if for some $\phi_0 >0$ and all $\beta$
satisfying $\|\beta_{S_0^c}\|_1 \le3 \|\beta_{S_0}\|_1$,
%
%e8.1 #&#
\begin{eqnarray}
\label{compat} \|\beta_{S_0}\|_1^2 \le
\beta^T \hat{\Sigma} \beta s_0/\phi_0^2,\quad
\hat{\Sigma} = n^{-1} \bx^T \bx.
\end{eqnarray}
Here $\beta_{A}$ denotes the components $\{\beta_j;j \in A\}$ where $A
\subseteq\{1,\ldots,p\}$. The number $\phi_0$ is called the compatibility
constant.

\emph{Aggregation of dependent $p$-values.}
Aggregation of dependent $p$-values can be generically done as follows.

%le1 #&#
\begin{lemm}[{[Implicitly contained in \citet{memepb09}]}]
Assume that we have\vadjust{\goodbreak} $B$ $p$-values $P^{(1)},\ldots
,P^{(B)}$ for testing a null-hypothesis $H_0$, that is, for every $b
\in\{1,\ldots
,B\}$ and any $0 < \alpha< 1$, $\PP_{H_0}[P^{(b)} \le\alpha] \le
\alpha$. Consider for any $0 < \gamma< 1$ the empirical $\gamma$-quantile
\begin{eqnarray*}
&&Q(\gamma) \\
&&\quad= \min \bigl(\mbox{empirical $\gamma$-quantile} \bigl
\{P^{(1)}/\gamma,\ldots,P^{(B)}/\gamma\bigr\},\\
&&\qquad 1 \bigr),
\end{eqnarray*}
and the minimum value of $Q(\gamma)$, suitably corrected with a factor,
over the range
$(\gamma_{\mathrm{min}},1)$ for some positive (small)
$0<\gamma_{\mathrm{min}} < 1$:
\begin{eqnarray*}
P = \min \Bigl(\bigl(1 - \log(\gamma_{\mathrm{min}})\bigr) \min
_{\gamma\in
(\gamma_{\mathrm{min}},1)} Q(\gamma), 1 \Bigr).
\end{eqnarray*}
Then, both $Q(\gamma)$ [for any fixed $\gamma\in(0,1)$] and $P$ are
conservative $p$-values satisfying for any $0 < \alpha< 1$,
$\PP_{H_0}[Q(\gamma) \le\alpha] \le
\alpha$ or $\PP_{H_0}[P \le\alpha] \le\alpha$, respectively.
\end{lemm}

\emph{Bounding the error of the estimated bias correction in the
desparsified Lasso.} We will argue now why the error from the bias
correction
\[
\sum_{k \neq j} \sqrt{n} P_{jk}\bigl(\hat{
\beta}_k - \beta^0_k\bigr)
\]
is negligible. From the KKT conditions when using the Lasso of $\bx^{(j)}$
versus $\bx^{(-j)}$, we have (B{\"u}hlmann\break and van~de Geer, \citeyear{pbvdg11}, cf. Lemma~2.1)
%
%e8.2 #&#
\begin{equation}
\label{KKT} \max_{k \neq j} 2 \bigl|n^{-1}
\bigl(X^{(k)}\bigr)^T Z^{(j)}\bigr| \le
\lambda_j.
\end{equation}
Therefore,
\begin{eqnarray*}
&&\biggl|\sqrt{n} \sum_{k \neq j} P_{jk}\bigl(\hat{
\beta}_k - \beta^0_k\bigr)\biggr| \\
&&\quad\le\sqrt{n}
\max_{k\neq j} |P_{jk}| \bigl\|\hat{\beta} -
\beta^0\bigr\|_1
\\
&&\quad\le2 \sqrt{n} \lambda_j\bigl \|\hat{\beta} - \beta^0
\bigr\|_1 \bigl(n^{-1} \bigl(\bx^{(j)}
\bigr)^T Z^{(j)}\bigr)^{-1}.
\end{eqnarray*}
Assuming sparsity and the compatibility condition (\ref{compat}), and when
choosing
$\lambda_j \asymp\sqrt{\log(p)/n}$, one can show that
$(n^{-1} (\bx^{(j)})^T Z^{(j)})^{-1} = O_P(1)$ and $\|\hat{\beta} -
\beta^0\|_1 = O_P(s_0 \sqrt{\log(p)/n})$ [for the latter, see
(\ref{lasso-ell1})]. Therefore,
\begin{eqnarray*}
&&\biggl|\sqrt{n} \sum_{k \neq j} P_{jk}\bigl(\hat{
\beta}_k - \beta^0_k\bigr)\biggr| \\
&&\quad\le
O_P\bigl(\sqrt{n} s_0 \sqrt{\log(p)/n}
\lambda_j\bigr) \\
&&\quad= O_P\bigl(s_0 \log(p)
n^{-1/2}\bigr),
\end{eqnarray*}
where the last bound follows by assuming $\lambda_j \asymp
\sqrt{\log(p)/n}$. Thus, if $s_0 \ll n^{1/2} / \log(p)$, the error from
bias correction is asymptotically negligible.

\emph{Choice of $\lambda_j$ for desparsified Lasso.}
We see from (\ref{KKT}) that the numerator of the error in the bias
correction term (i.e., the $P_{jk}$'s) is decreasing as $\lambda_j
\searrow
0$; for controlling the denominator, $\lambda_j$ should not be too
small to ensure that the denominator [i.e., $n^{-1} (\bx^{(j)})^T
Z^{(j)}$] behaves
reasonable (staying away from zero) for a fairly large range of
$\lambda_j$.

Therefore, the strategy is as follows:
\begin{longlist}[1.]
\item[1.] Compute a Lasso regression of $\bx^{(j)}$ versus all
other variables $\bx^{(-j)}$ using CV, and the corresponding residual
vector is
denoted by $Z^{(j)}$.
\item[2.] Compute $\|Z^{(j)}\|_2^2/((\bx^{(j)})^T Z^{(j)})^2$ which is the
asymptotic variance of $\hat{b}_j/\sigma_{\eps}$, assuming that the error
in the bias correction is negligible.
\item[3.] Increase the variance by 25\%, that is,
$V_j = 1.25 \|Z^{(j)}\|_2^2/((\bx^{(j)})^T Z^{(j)})^2$.
\item[4.] Search for the smallest $\lambda_j$ such that the corresponding
residual vector $Z^{(j)}(\lambda_j)$ satisfies
\begin{eqnarray*}
\bigl\|Z^{(j)}(\lambda_j)\bigr\|_2^2/\bigl(
\bigl(\bx^{(j)}\bigr)^T Z^{(j)}(
\lambda_j)\bigr)^2 \le V_j.
\end{eqnarray*}
\end{longlist}
This procedure is similar to the choice of $\lambda_j$ advocated in
\citet{zhangzhang11}.

\emph{Bounding the error of bias correction for the Ridge projection.}
The goal is to derive the formula (\ref{Ridge-repr}). Based on
(\ref{Ridge-distr}), we have
\begin{eqnarray*}
&&\sigma_{\eps}^{-1} \Omega_{R;jj}^{-1/2}
\bigl(\hat{b}_{R;j} - \beta^0_j\bigr)\\
&&\quad\approx
\Omega_{R;jj}^{-1/2} W_j / P_{R;jj} \\
&&\qquad{}+
\sigma_{\eps}^{-1} \Omega_{R;jj}^{-1/2}
\Delta_{R;j},\quad W \sim{\cal N}_p(0, \Omega_R),
\nonumber
\\
&&|\Delta_{R;j}|\le\max_{k \neq j} \biggl\llvert
\frac
{P_{R;jk}}{P_{R;jj}}\biggr\rrvert \bigl\|\hat{\beta} - \beta^0
\bigr\|_1.
\end{eqnarray*}

In relation to the result in Fact~\ref{th2} for the desparsified Lasso,
the problem here is that the behaviors of $\max_{k \neq j} |P_{R;jj}^{-1}
P_{R:jk}|$ and of the diagonal elements $\Omega_{R;jj}$ are hard to
control, but, fortunately, these quantities are fixed and observed for fixed
design $\bx$.

By invoking the compatibility constant for the design~$\bx$, we
obtain the bound for $\|\hat{\beta} - \beta^0\|_1 \le s_0 4\lambda/\phi_0$
in (\ref{lasso-ell1}) and, therefore, we can upper-bound
\[
|\Delta_{R;j}| \le4 s_0 \lambda/\phi_0^2
\max_{k \neq j} \biggl\llvert \frac{P_{R;jk}}{P_{R;jj}}\biggr\rrvert .\vadjust{\goodbreak}
\]
Asymptotically, for Gaussian errors, we have with high probability
%
%e8.3 #&#
\begin{eqnarray}
\label{delta-bound} |\Delta_{R;j}| &=& O\biggl(s_0 \sqrt{
\log(p)/n} \max_{k \neq
j}\biggl\llvert \frac{P_{R;jk}}{P_{R;jj}}\biggr
\rrvert \biggr)
\nonumber
\\[-8pt]
\\[-8pt]
\nonumber
&\le& O\biggl(\bigl(\log(p)/n\bigr)^{1/2
- \xi}\max_{k \neq j}
\biggl\llvert \frac{P_{R;jk}}{P_{R;jj}}\biggr\rrvert \biggr),
\end{eqnarray}
where the last inequality holds due to assuming $s_0 =O((n/\log(p))^{\xi})$
for some $0 < \xi< 1/2$.
In practice, we use the bound from (\ref{delta-bound}) in the form
\begin{eqnarray*}
\Delta_{R\mathrm{bound};j} := \max_{k \neq
j} \biggl\llvert
\frac{P_{R;jk}}{P_{R;jj}}\biggr\rrvert \bigl(\log(p)/n\bigr)^{1/2
- \xi},
\end{eqnarray*}
with the typical choice $\xi= 0.05$.

%s8.2 #&#
\subsection{Confidence Intervals for Multi Sample-Splitting}\label
{subsec.appmssplitci}
We construct confidence intervals that satisfy the duality with the $p$-values
from equation (\ref{aggreg}), and, thus, they are corrected already for
multiplicity:
\begin{eqnarray*}
&&\mbox{$(1-\alpha)$\% CI} \\
&&\quad= \mbox{Those values } c \mbox{ for which
the $p$-value }\geq\\
&&\qquad \alpha\mbox{ for testing the null hypothesis }
H_{0,j}:\beta=c,
\\
&&\quad=\mbox{Those } c \mbox{ for which the $p$-value resulting from}\\
&&\qquad\mbox{the $p$-value
aggregation procedure is} \geq\alpha,
\\
&&\quad= \{c | P_j \geq\alpha\},
\\
&&\quad= \Bigl\{c | (1-\log{\gamma_{\mathrm{min}}})\inf_{\gamma\in(\gamma_{\mathrm{min}},1)}
Q_j(\gamma) \geq\alpha\Bigr\},
\\
&&\quad= \bigl\{c | \forall\gamma\in(\gamma_{\mathrm{min}},1): (1-\log{
\gamma_{\mathrm{min}}}) Q_j(\gamma) \geq\alpha\bigr\},
\\
&&\quad= \bigl\{c | \forall\gamma\in(\gamma_{\mathrm{min}},1):\\
&&\qquad \min\bigl(1,\mathrm{emp.}\
\gamma\  \mathrm{quantile} \bigl(P_{\mathrm{corr};j}^{[b]}\bigr)/\gamma\bigr)\geq\\
&&\qquad
\alpha/(1-\log{\gamma_{\mathrm{min}}})\bigr\} ,
\\
&&\quad= \bigl\{c | \forall\gamma\in(\gamma_{\mathrm{min}},1):\\
&&\qquad \mathrm{emp.}\  \gamma\
\mathrm{quantile} \bigl(P_{\mathrm{corr};j}^{[b]}\bigr)/\gamma\geq\\
&&\qquad\alpha/(1-\log{
\gamma_{\mathrm{min}}})\bigr\} ,
\\
&&\quad = \biggl\{c | \forall\gamma\in(\gamma_{\mathrm{min}},1):\\
&&\qquad \mathrm{emp.}\ \gamma\ \mathrm{quantile} \bigl(P_{\mathrm{corr};j}^{[b]}\bigr) \geq\frac{\alpha
\gamma}{(1-\log{\gamma_{\mathrm{min}}})}\biggr
\}.
\end{eqnarray*}

We will use the notation $\gamma^{[b]}$ for the position of $P_{\mathrm{corr};j}^{[b]}$
in the ordering by increasing the value of the corrected $p$-values
$P_{\mathrm{corr};j}^{[i]}$,
divided by $B$.

We can now rewrite our former expression in a form explicitly using our
information from
every sample split
%%TODO rewrite in equation form!
\begin{eqnarray*}
&& \mbox{$(1-\alpha)$\% CI}
\\
&&\quad= \biggl\{c |\forall b =1,\ldots,B: \bigl(\gamma^{[b]} \leq
\gamma_{\mathrm{min}}\bigr)\\
&&\qquad{}\lor\biggl(P_{\mathrm{corr};j}^{[b]} \geq
\frac{\alpha\gamma^{[b]}}{(1-\log{\gamma_{\mathrm{min}}})}\biggr) \biggr\}
\\
% &= \{c | \forall b =1,\ldots,B: (\gamma^{[b]}
% \leq\gamma_{\mathrm{min}})\lor(c \in\mbox{ the }
% \left(1-\frac{\alpha\gamma^{[b]}}{(1-\log{\gamma_{\mathrm{min}}})|
%\hat{S}^{[b]}|}\right)100\%
% \mbox{CI for split b}) \},\\
&&\quad= \biggl\{c | \forall b
=1,\ldots,B: \bigl(\gamma^{[b]} \leq\gamma_{\mathrm{min}}\bigr)\\
&&\qquad{}\lor
\biggl(c \in\mbox{ the } \biggl(1-\frac{\alpha\gamma^{[b]}}{(1-\log{\gamma_{\mathrm{min}}})|\hat
{S}^{[b]}|} \biggr)\\
&&\qquad{}\cdot 100\%
\mbox{ CI for split $b$}\biggr) \biggr\}.
% &=\mbox{Those }c \mbox{ for which for all splits (excluding those
% with the } \gamma_{\mathrm{min}}B \mbox{ smallest pvalues), }\\
% &P_{corr;j}^{[b]} \geq\frac{0.05\gamma^{[b]}}{(1-\log{
%\gamma_{min}})},\\
% &= \mbox{Those }c \mbox{ for which for all splits (excluding those
% with the } \gamma_{min}B \mbox{ smallest pvalues), }\\
% & \mbox{the }c \mbox{ lie in the }
% \left(1-\frac{0.05\gamma^{[b]}}{(1-\log{\gamma_{min}})|
%\hat{S}^{[b]}|}\right)100\% \mbox{
% confidence intervals for the corresponding split.}\\
% &= \mbox{Those }c \mbox{ for which for all splits (excluding those
% with the } \gamma_{min}B \mbox{ smallest \emph{Standard errors}), }
%\\
% & \mbox{the }c \mbox{ lie in the }
% \left(1-\frac{0.05\gamma^{[b]}}{(1-\log{\gamma_{min}})|
%\hat{S}^{[b]}|}\right)100\% \mbox{
% confidence intervals fo the corresponding split.}
\end{eqnarray*}

For single testing (not adjusted for multiplicity), the corresponding
confidence interval becomes
\begin{eqnarray*}
& &\mbox{$(1-\alpha)$\% CI}
\\
&&\quad = \biggl\{c | \forall b =1,\ldots,B: \bigl(\gamma^{[b]} \leq
\gamma_{\mathrm{min}}\bigr)\\
&&\qquad{}\lor\biggl(c \in\mbox{ the } \biggl(1-
\frac{\alpha\gamma^{[b]}}{(1-\log{\gamma_{\mathrm{min}}})} \biggr)\\
&&\qquad{}\cdot 100\% \mbox{ CI for split $b$}\biggr) \biggr\}.
\end{eqnarray*}

If one has starting points with one being in the confidence interval
and the other one outside of it, one can apply the bisection method to
find the bound in between these points.

%s8.3 #&#
\subsection{Weighted Squared Error Approach for General
GLM}\label{subsec.app.general.wsqerr}
We describe the approach presented in Section~\ref{subsec.GLMweighted}
in a
more general way. One algorithm for fitting generalized linear models
is to calculate the
maximum likelihood estimates $\hat{\beta}$ by applying iterative weighted
least squares (\cite{mccullagh1989generalized}).

As in Section~\ref{subsec.GLMweighted}, the idea is now to apply a standard
l1-penalized fitting of the model, then build up the weighted least squares
problem at the l1-solution and apply our linear model methods on this
problem.

From \citet{mccullagh1989generalized}, using the notation $\hat{z}_i =
g^{-1}((\mathbf X \hat{\beta})_i), i=1 , \ldots, n$,
the adjusted response variable becomes
\begin{eqnarray}
Y_{i,\mathrm{adj}} = (\mathbf X \hat{\beta})_i + (Y_i-
\hat{z}_i) \frac
{\partial
g(z)}{\partial z} \bigg|_{z=\hat{z}_i},\nonumber\\
 \eqntext{i = 1 , \ldots, n .}
\end{eqnarray}

We then get a weighted least squares problem
\[
\hat{\beta}_{\mathrm{new}} = \argmin_{\beta} (Y_{\mathrm{adj}} - \mathbf
X \beta)^T \mathbf W (Y_{\mathrm{adj}} - \mathbf X \beta),
\]
with weights
\begin{eqnarray*}
&&\mathbf W^{-1} \\
&&\quad= %
\left(\matrix{\displaystyle \biggl(\frac{\partial g(z)}{\partial z}
\biggr)^2 \bigg|_{z=\hat{z}_1} V(\hat{z}_1) & 0 \vspace*{2pt}\cr
0 & \displaystyle\biggl(\frac{\partial g(z)}{\partial z}\biggr)^2
\bigg|_{z=\hat{z}_2} V(\hat{z}_2)  \vspace*{2pt}
\cr
\vdots& \ddots\vspace*{2pt}
\cr
0 & \ldots  }
\right.
\\
&&\quad\hspace*{20pt}\left.\matrix{\ldots& 0
\vspace*{2pt}\cr
\ddots& \vdots
\vspace*{2pt}\cr
\ddots& 0
\vspace*{2pt}\cr
0 & \displaystyle\biggl(
\frac{\partial g(z)}{\partial z}\biggr)^2 \bigg|_{z=\hat{z}_n} V(\hat{z}_n)}
\right),
\end{eqnarray*}
with variance function $V(z)$.

The variance function $V(z)$ is related to the variance of the response
$Y$. To more clearly define this relation, we assume that the response $Y$
has a distribution of the form described in \citet{mccullagh1989generalized}:
\[
f_Y(y;\theta,\phi) = \exp{\bigl[\bigl(y \theta- b(\theta)\bigr)/a(
\phi) + c(y,\phi)\bigr]},
\]
with known functions $a(\cdot)$, $b(\cdot)$ and $c(\cdot)$. $\theta$ is
the canonical parameter and $\phi$ is the dispersion parameter.

As defined in \citet{mccullagh1989generalized}, the variance function
is then
related to the variance of the response in the following way:
\[
\Var(Y) = b^{\prime\prime}(\theta)a(\phi)=V\bigl(g^{-1}\bigl(\mathbf X
\beta^0\bigr)\bigr) a(\phi).
\]

We rewrite $Y_{w} = \sqrt{\mathbf W} Y_{\mathrm{adj}}$ and $X_w =
\sqrt{\mathbf W} \mathbf X$ to get
\[
\hat{\beta}_{\mathrm{new}} = \argmin_{\beta} (Y_w - \mathbf
X_w \beta)^T(Y_w - \mathbf X_w
\beta).
\]

The linear model methods can now be applied to $Y_{w}$ and
$\mathbf X_{w}$, thereby the estimate $\hat{\sigma}_{\eps}$ has to
be set to the value
1.
\end{appendix}

% zodis "Acknowledgments" paliekamas pagal autoriu%
\section*{Acknowledgments}
We would like to thank some reviewers for
insightful and constructive comments.

\begin{supplement}[id=suppA]
\stitle{Supplement to ``High-Dimensional Inference:\break \mbox{Confidence} Intervals,
$p$-Values and \textsf{R}-Software \texttt{hdi}''}
\slink[doi]{10.1214/15-STS527SUPP} %[doi,text={...}] - jei reikiasuskaldyti doi
\sdatatype{.pdf}
\sfilename{sts527\_supp.pdf}
\sdescription{The supplemental article contains additional empirical
results.}
\end{supplement}
%
%\begin{supplement}[id=suppA]
%\slink[doi]{COMPLETED BY THE TYPESETTER}
%\end{supplement}

%\bibliographystyle{apalike}
%\bibliography{reference}

\begin{thebibliography}{64}
% pybtex-1.43. Style name=ims, version=2.92, label_style=nameyear, sorting_style=complex, cfg=None, language=None.


%b1 ###
%b1 #&#
\bibitem[\protect\citeauthoryear{Barber and Cand{\`e}s}{2015}]{foygcand14}
\begin{barticle}[mr]
\bauthor{\bsnm{Barber},~\bfnm{Rina~Foygel}\binits{R.~F.}} \AND
\bauthor{\bsnm{Cand{\`e}s},~\bfnm{Emmanuel~J.}\binits{E.~J.}}
(\byear{2015}).
\btitle{Controlling the false discovery rate via knockoffs}.
\bjournal{Ann. Statist.}
\bvolume{43}
\bpages{2055--2085}.
\bid{doi={10.1214/15-AOS1337}, issn={0090-5364}, mr={3375876}}
\bptnote{check volume, check pages, check year}%
\end{barticle}
%

\bptok{imsref}%
% NOT OUTPUTTED:
%   number = 5
%   doi = http://dx.doi.org/10.1214/15-AOS1337
%   fjournal = The Annals of Statistics
\endbibitem

%b2 ###
%b2 #&#
\bibitem[\protect\citeauthoryear{Belloni, Chernozhukov and Kato}{2015}]{beletal13}
\begin{barticle}[mr]
\bauthor{\bsnm{Belloni},~\bfnm{A.}\binits{A.}},
\bauthor{\bsnm{Chernozhukov},~\bfnm{V.}\binits{V.}} \AND
\bauthor{\bsnm{Kato},~\bfnm{K.}\binits{K.}}
(\byear{2015}).
\btitle{Uniform post-selection inference for least absolute deviation regression and other $Z$-estimation problems}.
\bjournal{Biometrika}
\bvolume{102}
\bpages{77--94}.
\bid{doi={10.1093/biomet/asu056}, issn={0006-3444}, mr={3335097}}
\end{barticle}
%

\bptok{imsref}%
% NOT OUTPUTTED:
%   number = 1
%   doi = http://dx.doi.org/10.1093/biomet/asu056
%   fjournal = Biometrika
\endbibitem

%b3 ###
%b3 #&#
\bibitem[\protect\citeauthoryear{Belloni, Chernozhukov and Wang}{2011}]{belloni2011square}
\begin{barticle}[mr]
\bauthor{\bsnm{Belloni},~\bfnm{A.}\binits{A.}},
\bauthor{\bsnm{Chernozhukov},~\bfnm{V.}\binits{V.}} \AND
\bauthor{\bsnm{Wang},~\bfnm{L.}\binits{L.}}
(\byear{2011}).
\btitle{Square-root Lasso: Pivotal recovery of sparse signals via conic programming}.
\bjournal{Biometrika}
\bvolume{98}
\bpages{791--806}.
\bid{doi={10.1093/biomet/asr043}, issn={0006-3444}, mr={2860324}}
\end{barticle}
%

\bptok{imsref}%
% NOT OUTPUTTED:
%   number = 4
%   doi = http://dx.doi.org/10.1093/biomet/asr043
%   coden = BIOKAX
%   fjournal = Biometrika
\endbibitem

%b4 ###
%b4 #&#
\bibitem[\protect\citeauthoryear{Belloni et~al.}{2012}]{belloni2012sparse}
\begin{barticle}[mr]
\bauthor{\bsnm{Belloni},~\bfnm{A.}\binits{A.}},
\bauthor{\bsnm{Chen},~\bfnm{D.}\binits{D.}},
\bauthor{\bsnm{Chernozhukov},~\bfnm{V.}\binits{V.}} \AND
\bauthor{\bsnm{Hansen},~\bfnm{C.}\binits{C.}}
(\byear{2012}).
\btitle{Sparse models and methods for optimal instruments with an application to eminent domain}.
\bjournal{Econometrica}
\bvolume{80}
\bpages{2369--2429}.
\bid{doi={10.3982/ECTA9626}, issn={0012-9682}, mr={3001131}}
\end{barticle}
%

\bptok{imsref}%
% NOT OUTPUTTED:
%   number = 6
%   doi = http://dx.doi.org/10.3982/ECTA9626
%   coden = ECMTA7
%   fjournal = Econometrica. Journal of the Econometric Society
\endbibitem

%b5 ###
%b5 #&#
\bibitem[\protect\citeauthoryear{Benjamini and Yekutieli}{2001}]{benyek01}
\begin{barticle}[mr]
\bauthor{\bsnm{Benjamini},~\bfnm{Yoav}\binits{Y.}} \AND
\bauthor{\bsnm{Yekutieli},~\bfnm{Daniel}\binits{D.}}
(\byear{2001}).
\btitle{The control of the false discovery rate in multiple testing under dependency}.
\bjournal{Ann. Statist.}
\bvolume{29}
\bpages{1165--1188}.
\bid{doi={10.1214/aos/1013699998}, issn={0090-5364}, mr={1869245}}
\end{barticle}
%

\bptok{imsref}%
% NOT OUTPUTTED:
%   number = 4
%   doi = http://dx.doi.org/10.1214/aos/1013699998
%   coden = ASTSC7
%   fjournal = The Annals of Statistics
\endbibitem

%b6 ###
%b6 #&#
\bibitem[\protect\citeauthoryear{Benjamini and Yekutieli}{2005}]{beye05}
\begin{barticle}[mr]
\bauthor{\bsnm{Benjamini},~\bfnm{Yoav}\binits{Y.}} \AND
\bauthor{\bsnm{Yekutieli},~\bfnm{Daniel}\binits{D.}}
(\byear{2005}).
\btitle{False discovery rate-adjusted multiple confidence intervals for selected parameters}.
\bjournal{J. Amer. Statist. Assoc.}
\bvolume{100}
\bpages{71--93}.
%\bnote{With comments and a rejoinder by the authors}.
\bid{doi={10.1198/016214504000001907}, issn={0162-1459}, mr={2156820}}
\bptnote{check related, check pages}%
\end{barticle}
%

\bptok{imsref}%
% NOT OUTPUTTED:
%   number = 469
%   doi = http://dx.doi.org/10.1198/016214504000001907
%   coden = JSTNAL
%   fjournal = Journal of the American Statistical Association
\endbibitem

%b7 ###
%b7 #&#
\bibitem[\protect\citeauthoryear{Berk et~al.}{2013}]{berketal13}
\begin{barticle}[mr]
\bauthor{\bsnm{Berk},~\bfnm{Richard}\binits{R.}},
\bauthor{\bsnm{Brown},~\bfnm{Lawrence}\binits{L.}},
\bauthor{\bsnm{Buja},~\bfnm{Andreas}\binits{A.}},
\bauthor{\bsnm{Zhang},~\bfnm{Kai}\binits{K.}} \AND
\bauthor{\bsnm{Zhao},~\bfnm{Linda}\binits{L.}}
(\byear{2013}).
\btitle{Valid post-selection inference}.
\bjournal{Ann. Statist.}
\bvolume{41}
\bpages{802--837}.
\bid{doi={10.1214/12-AOS1077}, issn={0090-5364}, mr={3099122}}
\end{barticle}
%

\bptok{imsref}%
% NOT OUTPUTTED:
%   number = 2
%   doi = http://dx.doi.org/10.1214/12-AOS1077
%   fjournal = The Annals of Statistics
\endbibitem

%b8 ###
%b8 #&#
\bibitem[\protect\citeauthoryear{Bogdan et~al.}{2013}]{bogdan13}
\begin{bmisc}[auto:parserefs-M02]
\bauthor{\bsnm{Bogdan},~\bfnm{M.}\binits{M.}},
\bauthor{\bsnm{van~den Berg},~\bfnm{E.}\binits{E.}},
\bauthor{\bsnm{Su},~\bfnm{W.}\binits{W.}} \AND
\bauthor{\bsnm{Cand\`es},~\bfnm{E.}\binits{E.}}
(\byear{2013}).
\bhowpublished{Statistical estimation and testing via the sorted l1 norm.
Preprint. Available at \arxivurl{arXiv:1310.1969}}.
\end{bmisc}
%

\bptok{imsref}%
\endbibitem

%b9 ###
%b9 #&#
\bibitem[\protect\citeauthoryear{Bogdan et~al.}{2014}]{bogdan14}
\begin{bmisc}[auto:parserefs-M02]
\bauthor{\bsnm{Bogdan},~\bfnm{M.}\binits{M.}},
\bauthor{\bsnm{van~den Berg},~\bfnm{E.}\binits{E.}},
\bauthor{\bsnm{Sabatti},~\bfnm{C.}\binits{C.}},
\bauthor{\bsnm{Su},~\bfnm{W.}\binits{W.}} \AND
\bauthor{\bsnm{Cand\`es},~\bfnm{E.}\binits{E.}}
(\byear{2014}).
\bhowpublished{SLOPE---adaptive variable selection
via convex optimization.
Preprint. Available at \arxivurl{arXiv:1407.3824}}.
\end{bmisc}
%

\bptok{imsref}%
\endbibitem

%b10 ###
%b10 #&#
\bibitem[\protect\citeauthoryear{Breiman}{1996a}]{brei96}
\begin{barticle}[auto:parserefs-M02]
\bauthor{\bsnm{Breiman},~\bfnm{L.}\binits{L.}}
(\byear{1996}a).
\btitle{Bagging predictors}.
\bjournal{Mach. Learn.}
\bvolume{24}
\bpages{123--140}.
\end{barticle}
%

\bptok{imsref}%
\endbibitem

%b11 ###
%b11 #&#
\bibitem[\protect\citeauthoryear{Breiman}{1996b}]{brei96b}
\begin{barticle}[mr]
\bauthor{\bsnm{Breiman},~\bfnm{Leo}\binits{L.}}
(\byear{1996}b).
\btitle{Heuristics of instability and stabilization in model selection}.
\bjournal{Ann. Statist.}
\bvolume{24}
\bpages{2350--2383}.
\bid{doi={10.1214/aos/1032181158}, issn={0090-5364}, mr={1425957}}
\end{barticle}
%

\bptok{imsref}%
% NOT OUTPUTTED:
%   number = 6
%   doi = http://dx.doi.org/10.1214/aos/1032181158
%   coden = ASTSC7
%   fjournal = The Annals of Statistics
\endbibitem

%b12 ###
%b12 #&#
\bibitem[\protect\citeauthoryear{B{\"u}hlmann}{2013}]{pb13}
\begin{barticle}[mr]
\bauthor{\bsnm{B{\"u}hlmann},~\bfnm{Peter}\binits{P.}}
(\byear{2013}).
\btitle{Statistical significance in high-dimensional linear models}.
\bjournal{Bernoulli}
\bvolume{19}
\bpages{1212--1242}.
\bid{doi={10.3150/12-BEJSP11}, issn={1350-7265}, mr={3102549}}
\end{barticle}
%

\bptok{imsref}%
% NOT OUTPUTTED:
%   number = 4
%   doi = http://dx.doi.org/10.3150/12-BEJSP11
%   fjournal = Bernoulli. Official Journal of the Bernoulli Society for Mathematical Statistics and Probability
\endbibitem

%b13 ###
%b13 #&#
\bibitem[\protect\citeauthoryear{B{\"{u}}hlmann, Kalisch and Meier}{2014}]{bumeka13}
\begin{barticle}[auto:parserefs-M02]
\bauthor{\bsnm{B{\"{u}}hlmann},~\bfnm{P.}\binits{P.}},
\bauthor{\bsnm{Kalisch},~\bfnm{M.}\binits{M.}} \AND
\bauthor{\bsnm{Meier},~\bfnm{L.}\binits{L.}}
(\byear{2014}).
\btitle{High-dimensional statistics with a view towards applications in biology}.
\bjournal{Annual Review of Statistics and Its Applications}
\bvolume{1}
\bpages{255--278}.
\end{barticle}
%

\bptok{imsref}%
\endbibitem

%b14 ###
%b14 #&#
\bibitem[\protect\citeauthoryear{B{\"u}hlmann and Mandozzi}{2014}]{pbmand13}
\begin{barticle}[mr]
\bauthor{\bsnm{B{\"u}hlmann},~\bfnm{Peter}\binits{P.}} \AND
\bauthor{\bsnm{Mandozzi},~\bfnm{Jacopo}\binits{J.}}
(\byear{2014}).
\btitle{High-dimensional variable screening and bias in subsequent inference, with an empirical comparison}.
\bjournal{Comput. Statist.}
\bvolume{29}
\bpages{407--430}.
\bid{doi={10.1007/s00180-013-0436-3}, issn={0943-4062}, mr={3261821}}
\end{barticle}
%

\bptok{imsref}%
% NOT OUTPUTTED:
%   number = 3-4
%   doi = http://dx.doi.org/10.1007/s00180-013-0436-3
%   fjournal = Computational Statistics
\endbibitem

%b15 ###
%b15 #&#
\bibitem[\protect\citeauthoryear{B{\"u}hlmann, Meier and van~de Geer}{2014}]{covtestpblmvdg14}
\begin{barticle}[mr]
\bauthor{\bsnm{B{\"u}hlmann},~\bfnm{Peter}\binits{P.}},
\bauthor{\bsnm{Meier},~\bfnm{Lukas}\binits{L.}} \AND
\bauthor{\bsnm{van~de Geer},~\bfnm{Sara}\binits{S.}}
(\byear{2014}).
\btitle{Discussion: ``{A} significance test for the Lasso''}.
\bjournal{Ann. Statist.}
\bvolume{42}
\bpages{469--477}.
\bid{doi={10.1214/13-AOS1175A}, issn={0090-5364}, mr={3210971}}
\end{barticle}
%

\bptok{imsref}%
% NOT OUTPUTTED:
%   number = 2
%   doi = http://dx.doi.org/10.1214/13-AOS1175A
%   fjournal = The Annals of Statistics
\endbibitem

%b16 ###
%b16 #&#
\bibitem[\protect\citeauthoryear{B{\"u}hlmann and van~de Geer}{2011}]{pbvdg11}
\begin{bbook}[mr]
\bauthor{\bsnm{B{\"u}hlmann},~\bfnm{Peter}\binits{P.}} \AND
\bauthor{\bsnm{van~de Geer},~\bfnm{Sara}\binits{S.}}
(\byear{2011}).
\btitle{Statistics for High-Dimensional Data: Methods, Theory and Applications}.
%\bseries{Springer Series in Statistics}.
\bpublisher{Springer},
\blocation{Heidelberg}.
\bid{doi={10.1007/978-3-642-20192-9}, mr={2807761}}
\end{bbook}
%

\bptok{imsref}%
% NOT OUTPUTTED:
%   doi = http://dx.doi.org/10.1007/978-3-642-20192-9
%   isbn = 978-3-642-20191-2
%   fpage = xviii+556
\endbibitem

%b17 ###
%b17 #&#
\bibitem[\protect\citeauthoryear{B{\"u}hlmann and van~de Geer}{2015}]{pbvdg15}
\begin{barticle}[mr]
\bauthor{\bsnm{B{\"u}hlmann},~\bfnm{Peter}\binits{P.}} \AND
\bauthor{\bsnm{van~de Geer},~\bfnm{Sara}\binits{S.}}
(\byear{2015}).
\btitle{High-dimensional inference in misspecified linear models}.
\bjournal{Electron. J. Stat.}
\bvolume{9}
\bpages{1449--1473}.
\bid{doi={10.1214/15-EJS1041}, issn={1935-7524}, mr={3367666}}
\end{barticle}
%

\bptok{imsref}%
% NOT OUTPUTTED:
%   doi = http://dx.doi.org/10.1214/15-EJS1041
%   fjournal = Electronic Journal of Statistics
\endbibitem

%b18 ###
%b18 #&#
\bibitem[\protect\citeauthoryear{Candes and Tao}{2006}]{candes2006near}
\begin{barticle}[mr]
\bauthor{\bsnm{Candes},~\bfnm{Emmanuel~J.}\binits{E.~J.}} \AND
\bauthor{\bsnm{Tao},~\bfnm{Terence}\binits{T.}}
(\byear{2006}).
\btitle{Near-optimal signal recovery from random projections: Universal encoding strategies?}
\bjournal{IEEE Trans. Inform. Theory}
\bvolume{52}
\bpages{5406--5425}.
\bid{doi={10.1109/TIT.2006.885507}, issn={0018-9448}, mr={2300700}}
\end{barticle}
%

\bptok{imsref}%
% NOT OUTPUTTED:
%   number = 12
%   doi = http://dx.doi.org/10.1109/TIT.2006.885507
%   coden = IETTAW
%   fjournal = Institute of Electrical and Electronics Engineers. Transactions on Information Theory
\endbibitem

%b19 ###
%b19 #&#
\bibitem[\protect\citeauthoryear{Chandrasekaran, Parrilo and Willsky}{2012}]{chandrasekaran2012}
\begin{barticle}[mr]
\bauthor{\bsnm{Chandrasekaran},~\bfnm{Venkat}\binits{V.}},
\bauthor{\bsnm{Parrilo},~\bfnm{Pablo~A.}\binits{P.~A.}} \AND
\bauthor{\bsnm{Willsky},~\bfnm{Alan~S.}\binits{A.~S.}}
(\byear{2012}).
\btitle{Latent variable graphical model selection via convex optimization}.
\bjournal{Ann. Statist.}
\bvolume{40}
\bpages{1935--1967}.
\bid{doi={10.1214/11-AOS949}, issn={0090-5364}, mr={3059067}}
\end{barticle}
%

\bptok{imsref}%
% NOT OUTPUTTED:
%   number = 4
%   doi = http://dx.doi.org/10.1214/11-AOS949
%   fjournal = The Annals of Statistics
\endbibitem

%b20 ###
%b20 #&#
\bibitem[\protect\citeauthoryear{Chatterjee and Lahiri}{2013}]{chatter13}
\begin{barticle}[mr]
\bauthor{\bsnm{Chatterjee},~\bfnm{A.}\binits{A.}} \AND
\bauthor{\bsnm{Lahiri},~\bfnm{S.~N.}\binits{S.~N.}}
(\byear{2013}).
\btitle{Rates of convergence of the adaptive LASSO estimators to the oracle distribution and higher order refinements by the bootstrap}.
\bjournal{Ann. Statist.}
\bvolume{41}
\bpages{1232--1259}.
\bid{doi={10.1214/13-AOS1106}, issn={0090-5364}, mr={3113809}}
\end{barticle}
%

\bptok{imsref}%
% NOT OUTPUTTED:
%   number = 3
%   doi = http://dx.doi.org/10.1214/13-AOS1106
%   fjournal = The Annals of Statistics
\endbibitem

%b21 ###
%b21 #&#
\bibitem[\protect\citeauthoryear{Dempster, Laird and Rubin}{1977}]{dempster1977maximum}
\begin{barticle}[mr]
\bauthor{\bsnm{Dempster},~\bfnm{A.~P.}\binits{A.~P.}},
\bauthor{\bsnm{Laird},~\bfnm{N.~M.}\binits{N.~M.}} \AND
\bauthor{\bsnm{Rubin},~\bfnm{D.~B.}\binits{D.~B.}}
(\byear{1977}).
\btitle{Maximum likelihood from incomplete data via the EM algorithm}.
\bjournal{J.~R. Stat. Soc. Ser. B. Stat. Methodol.}
\bvolume{39}
\bpages{1--38}.
%\bnote{With discussion}.
\bid{issn={0035-9246}, mr={0501537}}
\bptnote{check volume, check related}%
\end{barticle}
%

\bptok{imsref}%
% NOT OUTPUTTED:
%   number = 1
%   fjournal = Journal of the Royal Statistical Society. Series B. Methodological
\endbibitem


%b22 #&#
\bibitem[\protect\citeauthoryear{Dezeure et al.}{2015}]{supplement}
\begin{bmisc}[author]
\bauthor{\bsnm{Dezeure},~\bfnm{R.}\binits{R.}},
\bauthor{\bsnm{B{\"{u}}hlmann},~\bfnm{P.}\binits{P.}},
\bauthor{\bsnm{Meier},~\bfnm{L.}\binits{L.}} \AND
\bauthor{\bsnm{Meinshausen},~\bfnm{N.}\binits{N.}}
(\byear{2015}).
\bhowpublished{Supplement to ``High-Dimensional Inference:
Confidence Intervals, $p$-Values and \textsf{R}-Software \texttt{hdi}.''
DOI:\doiurl{10.1214/15-STS527SUPP}}.
\bptok{imsref}%
\end{bmisc}
\endbibitem
%
%%b22 ###
%\bibitem[\protect\citeauthoryear{Dezeure et~al.}{2015}]{}
%\begin{bbook}[auto:parserefs-M02]
%\bauthor{\bsnm{Dezeure},~\bfnm{R.}\binits{R.}},
%\bauthor{\bsnm{B{\"{u}}hlmann},~\bfnm{P.}\binits{P.}},
%\bauthor{\bsnm{Meier},~\bfnm{L.}\binits{L.}} \AND
%\bauthor{\bsnm{Meinshausen},~\bfnm{N.}\binits{N.}}
%(\byear{2015}).
%\btitle{Supplement to ``high-Dimensional Inference: Confidence Intervals, P-Values and R-Software Hdi''. DOI: COMPLETED bY tHE TYPESETTER},
%\blocation{\qq{}}.
%\end{bbook}
%%
%
%\bptok{imsref}%
%\endbibitem

%b23 ###
%b23 #&#
\bibitem[\protect\citeauthoryear{Fan and Li}{2001}]{fan2001variable}
\begin{barticle}[mr]
\bauthor{\bsnm{Fan},~\bfnm{Jianqing}\binits{J.}} \AND
\bauthor{\bsnm{Li},~\bfnm{Runze}\binits{R.}}
(\byear{2001}).
\btitle{Variable selection via nonconcave penalized likelihood and its oracle properties}.
\bjournal{J. Amer. Statist. Assoc.}
\bvolume{96}
\bpages{1348--1360}.
\bid{doi={10.1198/016214501753382273}, issn={0162-1459}, mr={1946581}}
\end{barticle}
%

\bptok{imsref}%
% NOT OUTPUTTED:
%   number = 456
%   doi = http://dx.doi.org/10.1198/016214501753382273
%   coden = JSTNAL
%   fjournal = Journal of the American Statistical Association
\endbibitem

%b24 ###
%b24 #&#
\bibitem[\protect\citeauthoryear{Fan and Lv}{2008}]{fanlv07}
\begin{barticle}[mr]
\bauthor{\bsnm{Fan},~\bfnm{Jianqing}\binits{J.}} \AND
\bauthor{\bsnm{Lv},~\bfnm{Jinchi}\binits{J.}}
(\byear{2008}).
\btitle{Sure independence screening for ultrahigh dimensional feature space}.
\bjournal{J. R. Stat. Soc. Ser. B. Stat. Methodol.}
\bvolume{70}
\bpages{849--911}.
\bid{doi={10.1111/j.1467-9868.2008.00674.x}, issn={1369-7412}, mr={2530322}}
\bptnote{check related}%
\end{barticle}
%

\bptok{imsref}%
% NOT OUTPUTTED:
%   number = 5
%   doi = http://dx.doi.org/10.1111/j.1467-9868.2008.00674.x
%   fjournal = Journal of the Royal Statistical Society. Series B. Statistical Methodology
\endbibitem

%b25 ###
%b25 #&#
\bibitem[\protect\citeauthoryear{Fan and Lv}{2010}]{fanlv10}
\begin{barticle}[mr]
\bauthor{\bsnm{Fan},~\bfnm{Jianqing}\binits{J.}} \AND
\bauthor{\bsnm{Lv},~\bfnm{Jinchi}\binits{J.}}
(\byear{2010}).
\btitle{A selective overview of variable selection in high dimensional feature space}.
\bjournal{Statist. Sinica}
\bvolume{20}
\bpages{101--148}.
\bid{issn={1017-0405}, mr={2640659}}
\end{barticle}
%

\bptok{imsref}%
% NOT OUTPUTTED:
%   number = 1
%   fjournal = Statistica Sinica
\endbibitem

%b26 ###
%b26 #&#
\bibitem[\protect\citeauthoryear{Fan, Xue and Zou}{2014}]{fan2014}
\begin{barticle}[mr]
\bauthor{\bsnm{Fan},~\bfnm{Jianqing}\binits{J.}},
\bauthor{\bsnm{Xue},~\bfnm{Lingzhou}\binits{L.}} \AND
\bauthor{\bsnm{Zou},~\bfnm{Hui}\binits{H.}}
(\byear{2014}).
\btitle{Strong oracle optimality of folded concave penalized estimation}.
\bjournal{Ann. Statist.}
\bvolume{42}
\bpages{819--849}.
\bid{doi={10.1214/13-AOS1198}, issn={0090-5364}, mr={3210988}}
\end{barticle}
%

\bptok{imsref}%
% NOT OUTPUTTED:
%   number = 3
%   doi = http://dx.doi.org/10.1214/13-AOS1198
%   fjournal = The Annals of Statistics
\endbibitem

%b27 ###
%b27 #&#
\bibitem[\protect\citeauthoryear{Fithian, Sun and Taylor}{2014}]{fithian14}
\begin{bmisc}[auto:parserefs-M02]
\bauthor{\bsnm{Fithian},~\bfnm{W.}\binits{W.}},
\bauthor{\bsnm{Sun},~\bfnm{D.}\binits{D.}} \AND
\bauthor{\bsnm{Taylor},~\bfnm{J.}\binits{J.}}
(\byear{2014}).
\bhowpublished{Optimal inference after model selection.
Preprint. Available at \arxivurl{arXiv:1410.2597}}.
\end{bmisc}
%

\bptok{imsref}%
\endbibitem

%b28 ###
%b28 #&#
\bibitem[\protect\citeauthoryear{Hartigan}{1975}]{hartigan1975clustering}
\begin{bbook}[mr]
\bauthor{\bsnm{Hartigan},~\bfnm{John~A.}\binits{J.~A.}}
(\byear{1975}).
\btitle{Clustering Algorithms}.
\bpublisher{Wiley},
\blocation{New York}.
%\bnote{Wiley Series in Probability and Mathematical Statistics}.
\bid{mr={0405726}}
\end{bbook}
%

\bptok{imsref}%
% NOT OUTPUTTED:
%   fpage = xiii+351
\endbibitem

%b29 ###
%b29 #&#
\bibitem[\protect\citeauthoryear{Hastie, Tibshirani and Friedman}{2009}]{hastetal09}
\begin{bbook}[mr]
\bauthor{\bsnm{Hastie},~\bfnm{Trevor}\binits{T.}},
\bauthor{\bsnm{Tibshirani},~\bfnm{Robert}\binits{R.}} \AND
\bauthor{\bsnm{Friedman},~\bfnm{Jerome}\binits{J.}}
(\byear{2009}).
\btitle{The Elements of Statistical Learning:
Data Mining, Inference, and Prediction},
\bedition{2nd} ed.
%\bseries{Springer Series in Statistics}.
\bpublisher{Springer},
\blocation{New York}.
\bid{doi={10.1007/978-0-387-84858-7}, mr={2722294}}
\end{bbook}
%

\bptok{imsref}%
% NOT OUTPUTTED:
%   doi = http://dx.doi.org/10.1007/978-0-387-84858-7
%   isbn = 978-0-387-84857-0
%   fpage = xxii+745
\endbibitem

%b30 ###
%b30 #&#
\bibitem[\protect\citeauthoryear{Javanmard and Montanari}{2014}]{jamo13b}
\begin{barticle}[mr]
\bauthor{\bsnm{Javanmard},~\bfnm{Adel}\binits{A.}} \AND
\bauthor{\bsnm{Montanari},~\bfnm{Andrea}\binits{A.}}
(\byear{2014}).
\btitle{Confidence intervals and hypothesis testing for high-dimensional regression}.
\bjournal{J.~Mach. Learn. Res.}
\bvolume{15}
\bpages{2869--2909}.
\bid{issn={1532-4435}, mr={3277152}}
\end{barticle}
%

\bptok{imsref}%
% NOT OUTPUTTED:
%   fjournal = Journal of Machine Learning Research (JMLR)
\endbibitem

%b31 ###
%b31 #&#
\bibitem[\protect\citeauthoryear{Knight and Fu}{2000}]{knfu00}
\begin{barticle}[mr]
\bauthor{\bsnm{Knight},~\bfnm{Keith}\binits{K.}} \AND
\bauthor{\bsnm{Fu},~\bfnm{Wenjiang}\binits{W.}}
(\byear{2000}).
\btitle{Asymptotics for Lasso-type estimators}.
\bjournal{Ann. Statist.}
\bvolume{28}
\bpages{1356--1378}.
\bid{doi={10.1214/aos/1015957397}, issn={0090-5364}, mr={1805787}}
\end{barticle}
%

\bptok{imsref}%
% NOT OUTPUTTED:
%   number = 5
%   doi = http://dx.doi.org/10.1214/aos/1015957397
%   coden = ASTSC7
%   fjournal = The Annals of Statistics
\endbibitem

%b32 ###
%b32 #&#
\bibitem[\protect\citeauthoryear{Lee et~al.}{2013}]{lee13}
\begin{bmisc}[auto:parserefs-M02]
\bauthor{\bsnm{Lee},~\bfnm{J.}\binits{J.}},
\bauthor{\bsnm{Sun},~\bfnm{D.}\binits{D.}},
\bauthor{\bsnm{Sun},~\bfnm{Y.}\binits{Y.}} \AND
\bauthor{\bsnm{Taylor},~\bfnm{J.}\binits{J.}}
(\byear{2013}).
\bhowpublished{Exact post-selection inference,
with application to the {L}asso.
Preprint. Available at \arxivurl{arXiv:1311.6238}}.
\end{bmisc}
%

\bptok{imsref}%
\endbibitem

%b33 ###
%b33 #&#
\bibitem[\protect\citeauthoryear{Leeb and P{\"o}tscher}{2003}]{leebpoetsch03}
\begin{barticle}[mr]
\bauthor{\bsnm{Leeb},~\bfnm{Hannes}\binits{H.}} \AND
\bauthor{\bsnm{P{\"o}tscher},~\bfnm{Benedikt~M.}\binits{B.~M.}}
(\byear{2003}).
\btitle{The finite-sample distribution of post-model-selection estimators and uniform versus nonuniform approximations}.
\bjournal{Econometric Theory}
\bvolume{19}
\bpages{100--142}.
\bid{doi={10.1017/S0266466603191050}, issn={0266-4666}, mr={1965844}}
\end{barticle}
%

\bptok{imsref}%
% NOT OUTPUTTED:
%   number = 1
%   doi = http://dx.doi.org/10.1017/S0266466603191050
%   fjournal = Econometric Theory
\endbibitem

%b34 ###
%b34 #&#
\bibitem[\protect\citeauthoryear{Liu and Yu}{2013}]{liuyu13}
\begin{barticle}[mr]
\bauthor{\bsnm{Liu},~\bfnm{Hanzhong}\binits{H.}} \AND
\bauthor{\bsnm{Yu},~\bfnm{Bin}\binits{B.}}
(\byear{2013}).
\btitle{Asymptotic properties of {L}asso${}+{}$m{LS} and {L}asso${}+{}${R}idge in sparse high-dimensional linear regression}.
\bjournal{Electron. J. Stat.}
\bvolume{7}
\bpages{3124--3169}.
\bid{issn={1935-7524}, mr={3151764}}
\end{barticle}
%

\bptok{imsref}%
% NOT OUTPUTTED:
%   fjournal = Electronic Journal of Statistics
\endbibitem

%b35 ###
%b35 #&#
\bibitem[\protect\citeauthoryear{Lockhart et~al.}{2014}]{covtest14}
\begin{barticle}[mr]
\bauthor{\bsnm{Lockhart},~\bfnm{Richard}\binits{R.}},
\bauthor{\bsnm{Taylor},~\bfnm{Jonathan}\binits{J.}},
\bauthor{\bsnm{Tibshirani},~\bfnm{Ryan~J.}\binits{R.~J.}} \AND
\bauthor{\bsnm{Tibshirani},~\bfnm{Robert}\binits{R.}}
(\byear{2014}).
\btitle{A significance test for the Lasso}.
\bjournal{Ann. Statist.}
\bvolume{42}
\bpages{413--468}.
\bid{doi={10.1214/13-AOS1175}, issn={0090-5364}, mr={3210970}}
\bptnote{check pages}%
\end{barticle}
%

\bptok{imsref}%
% NOT OUTPUTTED:
%   number = 2
%   doi = http://dx.doi.org/10.1214/13-AOS1175
%   fjournal = The Annals of Statistics
\endbibitem

%b36 ###
%b36 #&#
\bibitem[\protect\citeauthoryear{Mandozzi and B{\"{u}}hlmann}{2015}]{manbu13}
\begin{bmisc}[auto:parserefs-M02]
\bauthor{\bsnm{Mandozzi},~\bfnm{J.}\binits{J.}} \AND
\bauthor{\bsnm{B\"uhlmann},~\bfnm{P.}\binits{P.}}
(\byear{2015}).
\bhowpublished{Hierarchical testing in the high-dimensional
setting with correlated variables.
\textit{J. Amer. Statist. Assoc.}
To appear.
DOI:\doiurl{10.1080/01621459.2015.1007209}.
Available at \arxivurl{arXiv:1312.5556}}.
\end{bmisc}
%

\bptok{imsref}%
\endbibitem

%b37 ###
%b37 #&#
\bibitem[\protect\citeauthoryear{McCullagh and Nelder}{1983}]{mccullagh1989generalized}
\begin{bbook}[mr]
\bauthor{\bsnm{McCullagh},~\bfnm{P.}\binits{P.}} \AND
\bauthor{\bsnm{Nelder},~\bfnm{J.~A.}\binits{J.~A.}}
(\byear{1983}).
\btitle{Generalized Linear Models},
\bedition{2nd} ed.
%\bseries{Monographs on Statistics and Applied Probability}.
\bpublisher{Chapman \& Hall},
\blocation{London}.
\bid{doi={10.1007/978-1-4899-3244-0}, mr={0727836}}
\end{bbook}
%

\bptok{imsref}%
% NOT OUTPUTTED:
%   doi = http://dx.doi.org/10.1007/978-1-4899-3244-0
%   isbn = 0-412-23850-0
%   fpage = xiii+261
\endbibitem

%b38 ###
%b38 #&#
\bibitem[\protect\citeauthoryear{Meier, Meinshausen and Dezeure}{2014}]{hdipackage}
\begin{bmisc}[auto:parserefs-M02]
\bauthor{\bsnm{Meier},~\bfnm{L.}\binits{L.}},
\bauthor{\bsnm{Meinshausen},~\bfnm{N.}\binits{N.}} \AND
\bauthor{\bsnm{Dezeure},~\bfnm{R.}\binits{R.}}
(\byear{2014}).
\bhowpublished{hdi: High-Dimensional Inference.
R package version 0.1-2}.
\end{bmisc}
%

\bptok{imsref}%
\endbibitem

%b39 ###
%b39 #&#
\bibitem[\protect\citeauthoryear{Meinshausen}{2008}]{Meins08}
\begin{barticle}[mr]
\bauthor{\bsnm{Meinshausen},~\bfnm{Nicolai}\binits{N.}}
(\byear{2008}).
\btitle{Hierarchical testing of variable importance}.
\bjournal{Biometrika}
\bvolume{95}
\bpages{265--278}.
\bid{doi={10.1093/biomet/asn007}, issn={0006-3444}, mr={2521583}}
\end{barticle}
%

\bptok{imsref}%
% NOT OUTPUTTED:
%   number = 2
%   doi = http://dx.doi.org/10.1093/biomet/asn007
%   coden = BIOKAX
%   fjournal = Biometrika
\endbibitem

%b40 ###
%b40 #&#
\bibitem[\protect\citeauthoryear{Meinshausen}{2015}]{meins13}
\begin{barticle}[auto:parserefs-M02]
\bauthor{\bsnm{Meinshausen},~\bfnm{N.}\binits{N.}}
(\byear{2015}).
\btitle{Group-bound: Confidence intervals for groups of variables in sparse high-dimensional regression without assumptions on the design}.
\bjournal{J. R. Stat. Soc. Ser. B. Stat. Methodol.}
\bnote{To appear.
DOI:\doiurl{10.1111/rssb.12094}.
Available at \arxivurl{arXiv:1309.3489}}.
\end{barticle}
%

\bptok{imsref}%
\endbibitem

%b41 ###
%b41 #&#
\bibitem[\protect\citeauthoryear{Meinshausen and B{\"u}hlmann}{2006}]{mebu06}
\begin{barticle}[mr]
\bauthor{\bsnm{Meinshausen},~\bfnm{Nicolai}\binits{N.}} \AND
\bauthor{\bsnm{B{\"u}hlmann},~\bfnm{Peter}\binits{P.}}
(\byear{2006}).
\btitle{High-dimensional graphs and variable selection with the Lasso}.
\bjournal{Ann. Statist.}
\bvolume{34}
\bpages{1436--1462}.
\bid{doi={10.1214/009053606000000281}, issn={0090-5364}, mr={2278363}}
\end{barticle}
%

\bptok{imsref}%
% NOT OUTPUTTED:
%   number = 3
%   doi = http://dx.doi.org/10.1214/009053606000000281
%   coden = ASTSC7
%   fjournal = The Annals of Statistics
\endbibitem

%b42 ###
%b42 #&#
\bibitem[\protect\citeauthoryear{Meinshausen and B{\"u}hlmann}{2010}]{mebu10}
\begin{barticle}[mr]
\bauthor{\bsnm{Meinshausen},~\bfnm{Nicolai}\binits{N.}} \AND
\bauthor{\bsnm{B{\"u}hlmann},~\bfnm{Peter}\binits{P.}}
(\byear{2010}).
\btitle{Stability selection}.
\bjournal{J. R. Stat. Soc. Ser. B. Stat. Methodol.}
\bvolume{72}
\bpages{417--473}.
\bid{doi={10.1111/j.1467-9868.2010.00740.x}, issn={1369-7412}, mr={2758523}}
\bptnote{check related}%
\end{barticle}
%

\bptok{imsref}%
% NOT OUTPUTTED:
%   number = 4
%   doi = http://dx.doi.org/10.1111/j.1467-9868.2010.00740.x
%   fjournal = Journal of the Royal Statistical Society. Series B. Statistical Methodology
\endbibitem

%b43 ###
%b43 #&#
\bibitem[\protect\citeauthoryear{Meinshausen, Meier and B{\"u}hlmann}{2009}]{memepb09}
\begin{barticle}[mr]
\bauthor{\bsnm{Meinshausen},~\bfnm{Nicolai}\binits{N.}},
\bauthor{\bsnm{Meier},~\bfnm{Lukas}\binits{L.}} \AND
\bauthor{\bsnm{B{\"u}hlmann},~\bfnm{Peter}\binits{P.}}
(\byear{2009}).
\btitle{{$p$}-values for high-dimensional regression}.
\bjournal{J. Amer. Statist. Assoc.}
\bvolume{104}
\bpages{1671--1681}.
\bid{doi={10.1198/jasa.2009.tm08647}, issn={0162-1459}, mr={2750584}}
\end{barticle}
%

\bptok{imsref}%
% NOT OUTPUTTED:
%   number = 488
%   doi = http://dx.doi.org/10.1198/jasa.2009.tm08647
%   coden = JSTNAL
%   fjournal = Journal of the American Statistical Association
\endbibitem

%b44 ###
%b44 #&#
\bibitem[\protect\citeauthoryear{Pearl}{2000}]{pearl00}
\begin{bbook}[mr]
\bauthor{\bsnm{Pearl},~\bfnm{Judea}\binits{J.}}
(\byear{2000}).
\btitle{Causality: Models, Reasoning, and Inference}.
\bpublisher{Cambridge Univ. Press},
\blocation{Cambridge}.
\bid{mr={1744773}}
\end{bbook}
%

\bptok{imsref}%
% NOT OUTPUTTED:
%   isbn = 0-521-77362-8
%   fpage = xvi+384
\endbibitem

%b45 ###
%b45 #&#
\bibitem[\protect\citeauthoryear{Reid, Tibshirani and Friedman}{2013}]{reidtibsh13}
\begin{bmisc}[auto:parserefs-M02]
\bauthor{\bsnm{Reid},~\bfnm{S.}\binits{S.}},
\bauthor{\bsnm{Tibshirani},~\bfnm{R.}\binits{R.}} \AND
\bauthor{\bsnm{Friedman},~\bfnm{J.}\binits{J.}}
(\byear{2013}).
\bhowpublished{A study of error variance estimation in Lasso regression.
Preprint. Available at \arxivurl{arXiv:1311.5274}}.
\end{bmisc}
%

\bptok{imsref}%
\endbibitem

%b46 ###
%b46 #&#
\bibitem[\protect\citeauthoryear{Shah and Samworth}{2013}]{shah13}
\begin{barticle}[mr]
\bauthor{\bsnm{Shah},~\bfnm{Rajen~D.}\binits{R.~D.}} \AND
\bauthor{\bsnm{Samworth},~\bfnm{Richard~J.}\binits{R.~J.}}
(\byear{2013}).
\btitle{Variable selection with error control: Another look at stability selection}.
\bjournal{J. R. Stat. Soc. Ser. B. Stat. Methodol.}
\bvolume{75}
\bpages{55--80}.
\bid{doi={10.1111/j.1467-9868.2011.01034.x}, issn={1369-7412}, mr={3008271}}
\end{barticle}
%

\bptok{imsref}%
% NOT OUTPUTTED:
%   number = 1
%   doi = http://dx.doi.org/10.1111/j.1467-9868.2011.01034.x
%   fjournal = Journal of the Royal Statistical Society. Series B. Statistical Methodology
\endbibitem

%b47 ###
%b47 #&#
\bibitem[\protect\citeauthoryear{Shao and Deng}{2012}]{shadeng11}
\begin{barticle}[mr]
\bauthor{\bsnm{Shao},~\bfnm{Jun}\binits{J.}} \AND
\bauthor{\bsnm{Deng},~\bfnm{Xinwei}\binits{X.}}
(\byear{2012}).
\btitle{Estimation in high-dimensional linear models with deterministic design matrices}.
\bjournal{Ann. Statist.}
\bvolume{40}
\bpages{812--831}.
\bid{doi={10.1214/12-AOS982}, issn={0090-5364}, mr={2933667}}
\end{barticle}
%

\bptok{imsref}%
% NOT OUTPUTTED:
%   number = 2
%   doi = http://dx.doi.org/10.1214/12-AOS982
%   fjournal = The Annals of Statistics
\endbibitem

%b48 ###
%b48 #&#
\bibitem[\protect\citeauthoryear{Spirtes, Glymour and Scheines}{2000}]{sgs00}
\begin{bbook}[mr]
\bauthor{\bsnm{Spirtes},~\bfnm{Peter}\binits{P.}},
\bauthor{\bsnm{Glymour},~\bfnm{Clark}\binits{C.}} \AND
\bauthor{\bsnm{Scheines},~\bfnm{Richard}\binits{R.}}
(\byear{2000}).
\btitle{Causation, Prediction, and Search},
\bedition{2nd} ed.
%\bseries{Adaptive Computation and Machine Learning}.
\bpublisher{MIT Press},
\blocation{Cambridge, MA}.
%\bnote{With additional material by David Heckerman,
%Christopher Meek, Gregory F. Cooper and Thomas Richardson, A Bradford Book}.
\bid{mr={1815675}}
\end{bbook}
%

\bptok{imsref}%
% NOT OUTPUTTED:
%   isbn = 0-262-19440-6
%   fpage = xxii+543
\endbibitem

%b49 ###
%b49 #&#
\bibitem[\protect\citeauthoryear{Sun and Zhang}{2012}]{sunzhang11}
\begin{barticle}[mr]
\bauthor{\bsnm{Sun},~\bfnm{Tingni}\binits{T.}} \AND
\bauthor{\bsnm{Zhang},~\bfnm{Cun-Hui}\binits{C.-H.}}
(\byear{2012}).
\btitle{Scaled sparse linear regression}.
\bjournal{Biometrika}
\bvolume{99}
\bpages{879--898}.
\bid{doi={10.1093/biomet/ass043}, issn={0006-3444}, mr={2999166}}
\end{barticle}
%

\bptok{imsref}%
% NOT OUTPUTTED:
%   number = 4
%   doi = http://dx.doi.org/10.1093/biomet/ass043
%   fjournal = Biometrika
\endbibitem

%b50 ###
%b50 #&#
\bibitem[\protect\citeauthoryear{Taylor et~al.}{2014}]{taylor14}
\begin{bmisc}[auto:parserefs-M02]
\bauthor{\bsnm{Taylor},~\bfnm{J.}\binits{J.}},
\bauthor{\bsnm{Lockhart},~\bfnm{R.}\binits{R.}},
\bauthor{\bsnm{Tibshirani},~\bfnm{R.}\binits{R.}} \AND
\bauthor{\bsnm{Tibshirani},~\bfnm{R.}\binits{R.}}
(\byear{2014}).
\bhowpublished{Exact post-selection inference for forward
stepwise and least angle regression.
Preprint. Available at \arxivurl{arXiv:1401.3889}}.
\end{bmisc}
%

\bptok{imsref}%
\endbibitem

%b51 ###
%b51 #&#
\bibitem[\protect\citeauthoryear{Tibshirani}{1996}]{tibs96}
\begin{barticle}[mr]
\bauthor{\bsnm{Tibshirani},~\bfnm{Robert}\binits{R.}}
(\byear{1996}).
\btitle{Regression shrinkage and selection via the Lasso}.
\bjournal{J. R. Stat. Soc. Ser. B. Stat. Methodol.}
\bvolume{58}
\bpages{267--288}.
\bid{issn={0035-9246}, mr={1379242}}
\end{barticle}
%

\bptok{imsref}%
% NOT OUTPUTTED:
%   url = http://links.jstor.org/sici?sici=0035-9246(1996)58:1<267:RSASVT>2.0.CO;2-G&origin=MSN
%   number = 1
%   coden = JSTBAJ
%   fjournal = Journal of the Royal Statistical Society. Series B. Methodological
\endbibitem

%b52 ###
%b52 #&#
\bibitem[\protect\citeauthoryear{van~de Geer}{2007}]{vandeGeer:07a}
\begin{binproceedings}[auto:parserefs-M02]
\bauthor{\bsnm{van~de Geer},~\bfnm{S.}\binits{S.}}
(\byear{2007}).
\btitle{The deterministic {L}asso}.
In \bbooktitle{JSM Proceedings}
\bpages{140}.
\bpublisher{American Statistical Association},
\blocation{Alexandria, VA}.
\end{binproceedings}
%

\bptok{imsref}%
\endbibitem

%b53 ###
%b53 #&#
\bibitem[\protect\citeauthoryear{van~de Geer}{2014}]{vdg14}
\begin{bmisc}[auto:parserefs-M02]
\bauthor{\bsnm{van~de Geer},~\bfnm{S.}\binits{S.}}
(\byear{2014}).
\bhowpublished{Statistical theory for high-dimensional models.
Preprint. Available at \arxivurl{arXiv:1409.8557}}.
\end{bmisc}
%

\bptok{imsref}%
\endbibitem

%b54 ###
%b54 #&#
\bibitem[\protect\citeauthoryear{van~de Geer}{2015}]{vdg15}
\begin{bmisc}[auto:parserefs-M02]
\bauthor{\bsnm{van~de Geer},~\bfnm{S.}\binits{S.}}
(\byear{2015}).
\bhowpublished{$\chi^2$-confidence sets in high-dimensional regression.
Preprint. Available at \arxivurl{arXiv:1502.07131}}.
\end{bmisc}
%

\bptok{imsref}%
\endbibitem

%b55 ###
%b55 #&#
\bibitem[\protect\citeauthoryear{van~de Geer and B{\"u}hlmann}{2009}]{van2009conditions}
\begin{barticle}[mr]
\bauthor{\bsnm{van~de Geer},~\bfnm{Sara~A.}\binits{S.~A.}} \AND
\bauthor{\bsnm{B{\"u}hlmann},~\bfnm{Peter}\binits{P.}}
(\byear{2009}).
\btitle{On the conditions used to prove oracle results for the {L}asso}.
\bjournal{Electron. J. Stat.}
\bvolume{3}
\bpages{1360--1392}.
\bid{doi={10.1214/09-EJS506}, issn={1935-7524}, mr={2576316}}
\end{barticle}
%

\bptok{imsref}%
% NOT OUTPUTTED:
%   doi = http://dx.doi.org/10.1214/09-EJS506
%   fjournal = Electronic Journal of Statistics
\endbibitem

%b56 ###
%b56 #&#
\bibitem[\protect\citeauthoryear{van~de Geer, B{\"u}hlmann and Zhou}{2011}]{geer11}
\begin{barticle}[mr]
\bauthor{\bsnm{van~de Geer},~\bfnm{Sara}\binits{S.}},
\bauthor{\bsnm{B{\"u}hlmann},~\bfnm{Peter}\binits{P.}} \AND
\bauthor{\bsnm{Zhou},~\bfnm{Shuheng}\binits{S.}}
(\byear{2011}).
\btitle{The adaptive and the thresholded {L}asso for potentially misspecified models (and a lower bound for the {L}asso)}.
\bjournal{Electron. J. Stat.}
\bvolume{5}
\bpages{688--749}.
\bid{doi={10.1214/11-EJS624}, issn={1935-7524}, mr={2820636}}
\end{barticle}
%

\bptok{imsref}%
% NOT OUTPUTTED:
%   doi = http://dx.doi.org/10.1214/11-EJS624
%   fjournal = Electronic Journal of Statistics
\endbibitem

%b57 ###
%b57 #&#
\bibitem[\protect\citeauthoryear{van~de Geer et~al.}{2014}]{vdgetal13}
\begin{barticle}[mr]
\bauthor{\bsnm{van~de Geer},~\bfnm{Sara}\binits{S.}},
\bauthor{\bsnm{B{\"u}hlmann},~\bfnm{Peter}\binits{P.}},
\bauthor{\bsnm{Ritov},~\bfnm{Ya'acov}\binits{Y.}} \AND
\bauthor{\bsnm{Dezeure},~\bfnm{Ruben}\binits{R.}}
(\byear{2014}).
\btitle{On asymptotically optimal confidence regions and tests for high-dimensional models}.
\bjournal{Ann. Statist.}
\bvolume{42}
\bpages{1166--1202}.
\bid{doi={10.1214/14-AOS1221}, issn={0090-5364}, mr={3224285}}
\end{barticle}
%

\bptok{imsref}%
% NOT OUTPUTTED:
%   number = 3
%   doi = http://dx.doi.org/10.1214/14-AOS1221
%   fjournal = The Annals of Statistics
\endbibitem

%b58 ###
%b58 #&#
\bibitem[\protect\citeauthoryear{Wasserman}{2014}]{wass14}
\begin{barticle}[mr]
\bauthor{\bsnm{Wasserman},~\bfnm{Larry}\binits{L.}}
(\byear{2014}).
\btitle{Discussion: ``{A} significance test for the Lasso''}.
\bjournal{Ann. Statist.}
\bvolume{42}
\bpages{501--508}.
\bid{doi={10.1214/13-AOS1175E}, issn={0090-5364}, mr={3210975}}
\end{barticle}
%

\bptok{imsref}%
% NOT OUTPUTTED:
%   number = 2
%   doi = http://dx.doi.org/10.1214/13-AOS1175E
%   fjournal = The Annals of Statistics
\endbibitem

%b59 ###
%b59 #&#
\bibitem[\protect\citeauthoryear{Wasserman and Roeder}{2009}]{WR08}
\begin{barticle}[mr]
\bauthor{\bsnm{Wasserman},~\bfnm{Larry}\binits{L.}} \AND
\bauthor{\bsnm{Roeder},~\bfnm{Kathryn}\binits{K.}}
(\byear{2009}).
\btitle{High-dimensional variable selection}.
\bjournal{Ann. Statist.}
\bvolume{37}
\bpages{2178--2201}.
\bid{doi={10.1214/08-AOS646}, issn={0090-5364}, mr={2543689}}
\end{barticle}
%

\bptok{imsref}%
% NOT OUTPUTTED:
%   number = 5A
%   doi = http://dx.doi.org/10.1214/08-AOS646
%   coden = ASTSC7
%   fjournal = The Annals of Statistics
\endbibitem

%b60 ###
%b60 #&#
\bibitem[\protect\citeauthoryear{Yuan and Lin}{2006}]{yuan06}
\begin{barticle}[mr]
\bauthor{\bsnm{Yuan},~\bfnm{Ming}\binits{M.}} \AND
\bauthor{\bsnm{Lin},~\bfnm{Yi}\binits{Y.}}
(\byear{2006}).
\btitle{Model selection and estimation in regression with grouped variables}.
\bjournal{J. R. Stat. Soc. Ser. B. Stat. Methodol.}
\bvolume{68}
\bpages{49--67}.
\bid{doi={10.1111/j.1467-9868.2005.00532.x}, issn={1369-7412}, mr={2212574}}
\end{barticle}
%

\bptok{imsref}%
% NOT OUTPUTTED:
%   number = 1
%   doi = http://dx.doi.org/10.1111/j.1467-9868.2005.00532.x
%   fjournal = Journal of the Royal Statistical Society. Series B. Statistical Methodology
\endbibitem

%b61 ###
%b61 #&#
\bibitem[\protect\citeauthoryear{Zhang}{2010}]{zhang2010}
\begin{barticle}[mr]
\bauthor{\bsnm{Zhang},~\bfnm{Cun-Hui}\binits{C.-H.}}
(\byear{2010}).
\btitle{Nearly unbiased variable selection under minimax concave penalty}.
\bjournal{Ann. Statist.}
\bvolume{38}
\bpages{894--942}.
\bid{doi={10.1214/09-AOS729}, issn={0090-5364}, mr={2604701}}
\end{barticle}
%

\bptok{imsref}%
% NOT OUTPUTTED:
%   number = 2
%   doi = http://dx.doi.org/10.1214/09-AOS729
%   coden = ASTSC7
%   fjournal = The Annals of Statistics
\endbibitem

%b62 ###
%b62 #&#
\bibitem[\protect\citeauthoryear{Zhang and Zhang}{2014}]{zhangzhang11}
\begin{barticle}[mr]
\bauthor{\bsnm{Zhang},~\bfnm{Cun-Hui}\binits{C.-H.}} \AND
\bauthor{\bsnm{Zhang},~\bfnm{Stephanie~S.}\binits{S.~S.}}
(\byear{2014}).
\btitle{Confidence intervals for low dimensional parameters in high dimensional linear models}.
\bjournal{J. R. Stat. Soc. Ser. B. Stat. Methodol.}
\bvolume{76}
\bpages{217--242}.
\bid{doi={10.1111/rssb.12026}, issn={1369-7412}, mr={3153940}}
\end{barticle}
%

\bptok{imsref}%
% NOT OUTPUTTED:
%   number = 1
%   doi = http://dx.doi.org/10.1111/rssb.12026
%   fjournal = Journal of the Royal Statistical Society. Series B. Statistical Methodology
\endbibitem

%b63 ###
%b63 #&#
\bibitem[\protect\citeauthoryear{Zou}{2006}]{zou06}
\begin{barticle}[mr]
\bauthor{\bsnm{Zou},~\bfnm{Hui}\binits{H.}}
(\byear{2006}).
\btitle{The adaptive Lasso and its oracle properties}.
\bjournal{J.~Amer. Statist. Assoc.}
\bvolume{101}
\bpages{1418--1429}.
\bid{doi={10.1198/016214506000000735}, issn={0162-1459}, mr={2279469}}
\end{barticle}
%

\bptok{imsref}%
% NOT OUTPUTTED:
%   number = 476
%   doi = http://dx.doi.org/10.1198/016214506000000735
%   coden = JSTNAL
%   fjournal = Journal of the American Statistical Association
\endbibitem

%b64 ###
%b64 #&#
\bibitem[\protect\citeauthoryear{Zou and Hastie}{2005}]{zou2005regularization}
\begin{barticle}[mr]
\bauthor{\bsnm{Zou},~\bfnm{Hui}\binits{H.}} \AND
\bauthor{\bsnm{Hastie},~\bfnm{Trevor}\binits{T.}}
(\byear{2005}).
\btitle{Regularization and variable selection via the elastic net}.
\bjournal{J. R. Stat. Soc. Ser. B. Stat. Methodol.}
\bvolume{67}
\bpages{301--320}.
\bid{doi={10.1111/j.1467-9868.2005.00503.x}, issn={1369-7412}, mr={2137327}}
\end{barticle}
%

\bptok{imsref}%
% NOT OUTPUTTED:
%   number = 2
%   doi = http://dx.doi.org/10.1111/j.1467-9868.2005.00503.x
%   fjournal = Journal of the Royal Statistical Society. Series B. Statistical Methodology
\endbibitem
\end{thebibliography}
%
% imsref loaded by akundreckaite, 2015-09-04 12:27:06

%
%\begin{appendix}
%\section{}
%\end{appendix}

%\begin{thebibliography}{99}
%\bibitem[\protect\citeauthoryear{}{}]{r1}
%\bibitem{r1}
%\end{thebibliography}
\end{document}